\documentclass[aps,prd,onecolumn,showpacs,showkeys,amsmath,amssymb]{revtex4}
\usepackage{amsfonts}
\usepackage{amsmath}
\usepackage{graphicx}
\usepackage{subfigure}
\usepackage{dcolumn}
\usepackage{bm}
\usepackage{booktabs}
\usepackage{color}
\makeatletter
\newcommand{\figcaption}{\def\@captype{figure}\caption}
\newcommand{\tabcaption}{\def\@captype{table}\caption}

\newcommand{\Rmnum}[1]{\expandafter\@slowromancap\romannumeral #1@}
\def\hlinewd#1{%
  \noalign{\ifnum0=`}\fi\hrule \@height #1 \futurelet
   \reserved@a\@xhline}
\makeatother

\def\qq{\langle\bar qq\rangle}

\def\GGb{\langle g_s^2GG\rangle}

\def\GGGb{\langle g_s^3fGGG\rangle}
\def\qGqa{\langle\bar  qGq\rangle}
\def\qGqb{\langle\bar qg_s\sigma\cdot Gq\rangle}

\def\JJb{\langle g_s^4jj\rangle}

\def\f(s){\left[(\alpha+\beta)m_c^2-\alpha\beta s\right]}
\def\non{\\ \nonumber}

\usepackage{ulem}

\begin{document}

\title{Heavy tetraquark states and quarkonium hybrids}

\author{Wei Chen}
\email{wec053@mail.usask.ca} \affiliation{Department of Physics and
Engineering Physics, University of Saskatchewan, Saskatoon, SK, S7N
5E2, Canada}
\author{T. G. Steele}
\email{tom.steele@usask.ca} \affiliation{Department of Physics and
Engineering Physics, University of Saskatchewan, Saskatoon, SK, S7N
5E2, Canada}
\author{Shi-Lin Zhu}
\email{zhusl@pku.edu.cn} \affiliation{Department of Physics and
State Key Laboratory of Nuclear Physics and Technology, Peking
University, Beijing 100871, China, and Collaborative Innovation
Center of Quantum Matter, Beijing 100871, China}

\begin{abstract}
Many of the  $XYZ$  resonances observed by the
Belle, Babar, CLEO and BESIII collaborations in the past decade are difficult to interpret as  conventional quark-antiquark mesons, 
motivating the consideration of scenarios such as multi-quark states, meson molecules, and hybrids.   
After a brief introduction to QCD sum-rule methods, we provide a brief but comprehensive 
review of the mass spectra of the quarkonium-like 
tetraquark states $qQ\bar q\bar Q$, doubly charmed/bottomed
tetraquark states $QQ\bar q\bar q$ and the heavy quarkonium hybrid
states $\bar QGQ$ in the QCD sum rules approach.
Possible interpretations of the $XYZ$ resonances are briefly discussed.
\end{abstract}



\maketitle

 \section{Introduction}\label{sec:intro}
In the conventional quark model (QM), hadrons, including the $q\bar
q$ mesons and $qqq$ baryons, are bound states of quarks and
anti-quarks \cite{1985-Godfrey-p189-231,
2012-Beringer-p10001-10001}. The strong interaction of the colored
quarks and gluons emerges from the low-energy regime of quantum
chromodynamics (QCD), which is the most technically-challenging aspect of the
standard model (SM). The hadron spectrum is therefore of great
importance to our understanding of QCD.$\ast$

In the quark model, a $q\bar q$ meson is a color-singlet state,
which is composed of a quark color triplet and an antiquark color
antitriplet. For a neutral $q\bar q$ state, its parity and charge
conjugation parity are $P=(-)^{L+1}$ and $C=(-)^{L+S}$ respectively,
where $L$ is the relative momentum and $S$ is the total spin. Thus
the allowed $J^{PC}$ quantum numbers for a neutral $q\bar q$ meson
are $J^{PC}=0^{++}, 0^{-+}, 1^{++}, 1^{--}, 1^{+-}, 2^{++}, 2^{--},
2^{-+},\cdots$. Most of the experimentally observed mesons can be
interpreted as a conventional $q\bar q$ state. However, a $q\bar q$
pair can also be a color octet in QCD, which may combine with the
other colored objects such as an excited gluonic field to form a
hybrid meson ($\bar qgq$). 
Hybrid mesons are very interesting since
they are allowed to carry not only the ordinary quantum numbers of the 
$q\bar q$ mesons listed above, but also exotic quantum
numbers such as $J^{PC}=0^{--}, 0^{+-}, 1^{-+}, 2^{+-},\cdots$,
which are not accessible to simple $q\bar q$ states. The study of
the hybrid mesons provides an important platform to understand QCD
as the theory of the strong interaction, including characteristics of color confinement.

Since the first investigation of hybrid mesons by Jaffe in 1976
\cite{1976-Jaffe-p201-201}, light hybrids were extensively
studied in the MIT Bag model
\cite{1983-Barnes-p241-241,1983-Chanowitz-p211-211}, flux tube model
\cite{1985-Isgur-p869-869, 1995-Close-p1706-1709,
1995-Barnes-p5242-5256, 1999-Page-p34016-34016,
1995-Close-p233-254}, lattice QCD \cite{1999-McNeile-p264-266,
1999-Lacock-p261-263, 2003-Bernard-p74505-74505,
2005-Hedditch-p114507-114507} and QCD sum rules
\cite{1983-Govaerts-p262-262,
1984-Govaerts-p1-1,1984-Latorre-p169-169, 1987-Latorre-p347-347,
1986-Balitsky-p265-273, 2003-Jin-p14025-14025,
2000-Chetyrkin-p145-150, 1999-Zhu-p97502-97502}. To date, there has
been some experimental evidence of the exotic light hybrid with
$J^{PC}=1^{-+}$
\cite{1997-Thompson-p1630-1633,1998-Abele-p175-184,1999-Abele-p349-355,
2007-Adams-p27-31,2007-Klempt-p1-202, 2012-Ketzer-p25-25}. For the
heavy quarkonium hybrids, there are also many calculations performed
in the constituent gluon model \cite{1978-Horn-p898-898}, the flux
tube model \cite{1995-Barnes-p5242-5256}, QCD sum
rules \cite{1985-Govaerts-p215-215,1985-Govaerts-p575-575,
1987-Govaerts-p674-674,1999-Zhu-p31501-31501,2012-Qiao-p15005-15005,
2012-Harnett-p125003-125003, 2012-Berg-p34002-34002,
2013-Chen-p19-19,2014-Chen-p25003-25003}, nonrelativistic
QCD~\cite{1998-Chiladze-p34013-34013} and lattice
QCD~\cite{1990-Perantonis-p854-868, 1999-Juge-p4400-4403,
2006-Liu-p54510-54510, 2006-Luo-p34502-34502, 2011-Liu-p140-140,
2012-Liu-p126-126}.

The diquark pair is another interesting color non-singlet bilinear
operator in QCD. The color structure of a $qq$ diquark can be
symmetric $\mathbf{6_c}$ or antisymmetric $\mathbf{\bar 3_c}$ with similar 
flavor structure. Without inserting the covariant
derivative, the spin and parity of a diquark operator can be determined by
its color, flavor and Lorentz structures. The Lorentz structures are
classified by using different kinds of $\gamma$ matrices resulting in 
six distinct diquark operators in Lorentz space: $q^T_a Cq_b$,
$q^T_a C\gamma_5q_b$, $q^T_aC\gamma_\mu q_b$,
$q^T_aC\gamma_\mu\gamma_5q_b$, $q^T_a C\sigma_{\mu\nu}q_b$, and
$q^T_aC\sigma_{\mu\nu}\gamma_5q_b$, where $a, b$ are the color
indices. Since $q^T_aC\sigma_{\mu\nu}\gamma_5q_b$ and
$q^T_aC\sigma_{\mu\nu}q_b$ carry opposite parity, we consider both
operators although they overlap in the quantum numbers they can probe (see Table~\ref{table1}). 
These six diquark operators 
are listed in Table \ref{table1} with their spins, parities, flavor,
color and Lorentz structures, which are constrained by fermi
statistics. As will be seen below, these operators play a very important role in the study of the
tetraquark states.

Tetraquarks ($qq\bar q\bar q$) are composed of diquarks and
antidiquarks. 
They are bound by the color force between quarks. The
tetraquarks are generally unstable because they can decay easily into two
mesons through kinematically-allowed fall-apart mechanisms and thus they are expected to be very broad resonances.
The low-lying scalar mesons below 1 GeV have been considered as good
candidates of the light tetraquark states
\cite{2007-Chen-p369-372,2007-Chen-p94025-94025}. In the heavy quark
sector, some of the recently observed quarkoniumlike states were
suggested to be candidates for hidden charm/bottom $Q{\bar
Q}q\bar{q}$-type tetraquark states
\cite{2007-Matheus-p14005-14005,2007-Maiani-p182003-182003,
2006-Ebert-p214-219,2011-Chen-p34010-34010,2010-Chen-p105018-105018,2013-Du-p33104-33104},
where $Q$ denotes a heavy quark (bottom or charm) and $q$ denotes a
light quark (up, down, strange).

\begin{center}
\renewcommand{\arraystretch}{1.5}
\begin{tabular*}{11cm}{l|c|c|r}
\hlinewd{.8pt}
$q\Gamma q$                         & $J^P$             & ~~  States   ~~   & (Flavor, Color) \\
\hline $q_a^TC\gamma_5q_b$                 & $0^+$             &
$^1S_0$     &$(\mathbf{6_f},\mathbf{6_c}),
                                                               (\mathbf{\bar 3_f},\mathbf{\bar 3_c})$ \\
$q_a^TCq_b$                         & $0^-$             & $^3P_0$
&$(\mathbf{6_f},\mathbf{6_c}),
                                                               (\mathbf{\bar 3_f},\mathbf{\bar 3_c})$ \\
$q_a^TC\gamma_{\mu}\gamma_5q_b$     & $1^-$             & $^3P_1$
&$(\mathbf{6_f},\mathbf{6_c}),
                                                               (\mathbf{\bar 3_f},\mathbf{\bar 3_c})$ \\
$q_a^TC\gamma_{\mu}q_b$             & $1^+$             & $^3S_1$
&$(\mathbf{6_f},\mathbf{\bar 3_c}),
                                                               (\mathbf{\bar 3_f},\mathbf{6_c})$ \\
$q_a^TC\sigma_{\mu\nu}q_b$          &$\begin{cases}
                                     1^-,&\mbox{for}~\mu,\nu=1,2,3\\
                                     1^+,&\mbox{for}~\mu=0,\nu=1,2,3
                                     \end{cases}$       &$\begin{matrix}
                                                         ^1P_1\\^3S_1
                                                         \end{matrix}$   &$(\mathbf{6_f},\mathbf{\bar 3_c}),
                                                               (\mathbf{\bar 3_f},\mathbf{6_c})$ \\
$q_a^TC\sigma_{\mu\nu}\gamma_5q_b$  &$\begin{cases}
                                     1^+,&\mbox{for}~\mu,\nu=1,2,3\\
                                     1^-,&\mbox{for}~\mu=0,\nu=1,2,3
                                     \end{cases}$         &$\begin{matrix}
                                                         ^3S_1\\^1P_1
                                                         \end{matrix}$   &$(\mathbf{6_f},\mathbf{\bar 3_c}),
                                                               (\mathbf{\bar 3_f},\mathbf{6_c})$ \\
\hlinewd{.8pt}
\end{tabular*}
\tabcaption{Properties of the diquark operators.} \label{table1}
\end{center}

Another possibility for heavy tetraquarks are the doubly
charmed/bottomed $QQ\bar q\bar{q}$-type states. This color-singlet
four-quark configuration is also allowed in QCD. When the heavy
quark $QQ$ pair is spatially close, it would act as a pointlike
antiheavy quark color source $\bar Q$ and pick up two light
antiquarks $\bar q\bar{q}$ to form the bound state $QQ\bar
q\bar{q}$. Such a doubly charmed/bottomed tetraquark system is the
QCD analogue of the hydrogen molecule in QED, in which two electrons
are shared by two protons. The existence and stability of the doubly
charmed/bottomed tetraquark systems have been studied in many
different models, such as the MIT bag model \cite{1988-Carlson-p744-744}, 
QCD sum rules\cite{2013-Du-p14003-14003,2013-Chen-p-b,2012-Albuquerque-p492-498,
2007-Navarra-p166-172}, chiral quark model\cite{2008-Zhang-p437-440,1997-Pepin-p119-123,
2007-Ebert-p114015-114015}, constituent quark
model \cite{2006-Vijande-p54018-54018,2009-Vijande-p74010-74010,
1998-Brink-p6778-6787,1993-Silvestre-Brac-p457-470,1986-Zouzou-p457-457},
chiral perturbation theory \cite{1993-Manohar-p17-33} and some other
methods
\cite{2007-Cui-p7-13,2003-Gelman-p296-304,1996-Moinester-p349-362,
1994-Bander-p5478-5480,1982-Ader-p2370-2370,1991-Richard-p254-257,
1986-Lipkin-p242-242,1973-Lipkin-p267-271,2011-Carames-p291-295}.

In the past decade many unexpected mesons, the so called $XYZ$ states, were discovered at
B-factories. These states
contain a heavy quark-antiquark pair and are above the
open-charm/bottom threshold. To date, there are 15 neutral and 5
charged states in the charmonium sector while one neutral and two
charged states in the bottomonium sector \cite{2013-Bodwin-p-}. Some
of these states are surprisingly narrow. Some are even charged. Many
of these states do not fit in the conventional quark model and are
considered as candidates for exotic states. The underlying
structure of these newly observed $XYZ$ states inspired the
extensive study of the hadron spectrum. Many theoretical
speculations have been proposed to interpret these new resonances,
such as molecular states, quarkoniumlike tetraquark states,
quarkonium hybrids and conventional quarkonium states. To understand
the nature of these $XYZ$ states, further theoretical investigations
of the exotic hadron spectrum are still needed.

We organize this review as follows. In Sec.~\Rmnum{2}, we briefly
introduce the general procedure of QCD sum rules, including the
two-point correlation functions, the operator product expansion, the
Borel transform, quark-hadron duality and the mass sum rules. In
Sec.~\Rmnum{3}, we give the interpolating currents of the
quarkoniumlike tetraquark systems, the doubly charmed/bottomonium
tetraquark systems and the quarkonium hybrid systems. The
correlation functions and spectral densities are calculated using
these currents. We perform QCD sum rule analysis {of all three hadron
systems and extract the masses of the lowest lying states. In the
last section we summarize our results and comment on their
implications for heavy quarkonium spectroscopy.

 \section{QCD sum rules}\label{sec:QSR}
QCD sum rules provide a very powerful nonperturbative method for studying 
hadron structures \cite{1979-Shifman-p385-447,
1985-Reinders-p1-1,2000-Colangelo-p1495-1576,2010-Nielsen-p41-83}.
IN addition to the operator product expansion, a key idea of the method is 
quark-hadron duality: the equivalence of (integrated) 
correlation functions at both the hadronic level and the
quark-gluonic level.

 \subsection{Two-point correlation function}
In general, the two-point correlation function for a scalar or
pseudoscalar operator is defined as
\begin{equation}
\Pi(q)= i \int d^4xe^{iq\cdot x}\langle
0|T[J(x)J^{\dag}(0)]|0\rangle, \label{Pi_scalar}
\end{equation}
where $J(x)$ is an interpolating current that can couple to a scalar
or pseudoscalar resonance and $T$ denotes the time-ordered product. For
a vector or axial-vector interpolating current $J_{\mu}(x)$, the
corresponding two-point correlation function reads
\begin{equation}
\begin{split}
\Pi_{\mu\nu}(q)&=i\int d^4x e^{iq\cdot x}\langle 0|T [J_\mu(x)J_\nu^\dagger(0)]|0\rangle\\
&=\eta_{\mu\nu}\Pi_1(q^2)+\frac{q_\mu q_\nu}{q^2}\Pi_0(q^2),
\label{Pi_vector}
\end{split}
\end{equation}
in which $\eta_{\mu\nu}=q_\mu q_\nu/q^2-g_{\mu\nu}$ is the tensor
structure of the spin-1 invariant function $\Pi_1(q^2)$. The spin-0
invariant function $\Pi_0(q^2)$ also appears in 
Eq.~\eqref{Pi_vector} when $J_\mu(x)$ is not a conserved vector current.
We will introduce the choice of the interpolating currents that
couple to the heavy tetraquarks and quarkonium hybrids in the next
section.

At the hadronic level, the invariant function $\Pi(q^2)$ can be
expressed in the form of the dispersion relation with its imaginary
part
\begin{eqnarray}
\Pi(q^2)=\frac{(q^2)^N}{\pi}\int\frac{\mbox{Im}\Pi(s)}{s^N(s-q^2-i\epsilon)}ds+\sum_{n=0}^{N-1}b_n(q^2)^n,
\label{dispersionrelation}
\end{eqnarray}
in which $b_n$ are the $N$ unknown subtraction constants which can
be removed by taking the Borel transform. With a narrow resonance
approximation, the imaginary part of the correlation function is
obtained by inserting intermediate states $|n\rangle$ for the hadron
we want to study. The imaginary part can be written as a sum over
$\delta$ functions,
\begin{eqnarray}
\text{Im}\Pi(s)=\pi\sum_n\delta(s-m_n^2)\langle0|J|n\rangle\langle
n|J^{\dagger}|0\rangle,  \label{Imaginary}
\end{eqnarray}
in which the intermediate states $|n\rangle$ carry the same quantum
numbers as the interpolating current $J(x)$. The correlation
function $\Pi(q^2)$ contains the contributions from all resonances
that can couple to $J(x)$, including the lowest lying ground state
and the excited higher states. In QCD sum rules, one usually
parametrizes the spectral function $\rho(s)$ with a pole plus
continuum approximation,
\begin{eqnarray}
\rho(s)=\frac{1}{\pi}\text{Im}\Pi(s)=f_X^2\delta(s-m_X^2)+
\mbox{continuum},  \label{Phenrho}
\end{eqnarray}
where $m_X$ is the mass of the lowest lying resonance $|X\rangle$
and $f_X$ is the coupling parameter of the current $J(x)$ with
$|X\rangle$. For a scalar or pseudoscalar current $J(x)$ and vector
or axial-vector current $J_{\mu}(x)$, we have
\begin{eqnarray}
\langle0|J|X\rangle&=&f_X, \label{coupling parameter1}
\\
\langle0|J_{\mu}|X\rangle&=&f_X\epsilon_{\mu}, \label{coupling
parameter2}
\end{eqnarray}
where $\epsilon_{\mu}$ is the polarization vector ($\epsilon\cdot
q=0$).

 \subsection{Operator product expansion}
The two-point correlation function can also be calculated at the
quark-gluonic level via the operator product expansion (OPE)
\cite{1969-Wilson-p1499-1512}:
\begin{equation}
\Pi(q)= i \int d^4xe^{iq\cdot x}\langle
0|T[J(x)J^{\dag}(0)]|0\rangle=\sum_nC_n(Q^2)O_n\,,~Q^2=-q^2\,,
\end{equation}
where $C_n(Q^2)$ are the Wilson coefficients and $O_n$
are the vacuum expectation values of the local gauge invariant
operators constructed from the quark and gluon fields. The Wilson
coefficients $C_n(Q^2)$ can be calculated in perturbation theory and
expressed in terms of the QCD parameters such as the quark mass and
the strong coupling constant $\alpha_s$. The long distance
nonperturbative effects are included in the various condensates
$O_n$, which are ordered by increasing dimension in the expansion.
Except for the unit operator $I$, the QCD vacuum condensates up to dimension-eight include:
the quark condensate $\qq$, gluon condensate $\GGb$, quark-gluon
mixed condensate $\qGqb$, tri-gluon condensate $\GGGb$, four quark
condensate $\qq^2$, and condensate $\qq\qGqb$.

To calculate the correlation function $\Pi(q)$, we need the full
quark propagator including quark and gluon condensates \cite{1985-Reinders-p1-1,1993-Yang-p3001-3012}
\begin{eqnarray}
\nonumber iS^{ab}_q(x) &=&  \frac{i\delta^{ab}}{2\pi^2x^4}\hat{x}
+\frac{i}{32\pi^2}\frac{\lambda^n_{ab}}{2}g_sG_{\mu\nu}^n\frac{1}
{x^2}(\sigma^{\mu\nu}\hat{x}+\hat{x}\sigma^{\mu\nu})-\frac{\delta^{ab}}{12}\langle\bar{q}q\rangle+
\frac{\delta^{ab}x^2}{192}\qGqa-\frac{m_q\delta^{ab}}{4\pi^2x^2}
\\&&+\frac{i\delta^{ab}m_q\langle\bar{q}q\rangle}{48}\hat{x}
-\frac{im_q\qGqa\delta^{ab}x^2\hat{x}}{1152}, \label{prop_light}
\\
iS^{ab}_Q(p) &=&
\frac{i\delta^{ab}}{\hat{p}-m_Q}+\frac{i}{4}g_s\frac{\lambda^n_{ab}}{2}G_{\mu\nu}^n
\frac{\sigma^{\mu\nu}(\hat{p}+m_Q)+(\hat{p}+m_Q)\sigma^{\mu\nu}}
{(p^2-m_Q^2)^2}+\frac{i\delta^{ab}}{12}\GGb
m_Q\frac{p^2+m_Q\hat{p}}{(p^2-m_Q^2)^4},
\end{eqnarray}
in which $q$ represents $u$, $d$ or $s$ quarks and $Q$ represents $c$
or $b$ quarks. The superscripts $a, b$ are the color indices and
$\hat x=\gamma_{\mu}x^{\mu}$, $\hat p=\gamma_{\mu}p^{\mu}$. One
notes that for the light quarks, we use the propagator in 
coordinate space. For heavy quarks, the momentum space
expression is sometimes more convenient.

 \subsection{Mass sum rules}
 The fundamental assumption in the QCD sum rules is quark-hadron duality, which matches the two
 descriptions of the correlation function at both the hadronic level and quark-gluon level. To pick out the
 lowest lying resonance of interest, one defines the Borel transform,
 \begin{eqnarray}
 B\left[\Pi(q^2)\right]=\lim_{{-q^2, n\to\infty\atop -q^2/n\equiv M_B^2}}\frac{1}{n!}(-q^2)^{n+1}\left(\frac{d}{dq^2}\right)^n\Pi(q^2),
 \end{eqnarray}
where the Borel mass $M_B^2\equiv -q^2/n$ is introduced instead of
$q^2$. Performing the Borel transform of the dispersion relation in
Eq.~\eqref{dispersionrelation}, we can remove the unknown
subtraction terms and suppress the contributions from the excited
states and continuum. On the OPE side, the Borel transform can
improve the convergence of the OPE series by suppressing the
contribution from the high dimension condensates.

To establish the mass sum rules, we perform the Borel transform of 
the correlation function $\Pi(q^2)$ obtained at both levels
\begin{eqnarray}
f_X^2m_X^{2k}e^{-m_X^2/M_B^2}=\int_{4m_Q^2}^{s_0}dse^{-s/M_B^2}\rho(s)s^k=\mathcal{L}_{k}\left(s_0,
M_B^2\right), \label{sumrule}
\end{eqnarray}
where $s_0$ is the continuum threshold. On the left hand of 
Eq.~\eqref{sumrule}, we have used the spectral function defined in 
Eq.~\eqref{Phenrho}. The lowest lying hadron mass is then extracted as,
\begin{eqnarray}
m_X\left(s_0, M_B^2\right)=\sqrt{\frac{\mathcal{L}_{1}\left(s_0,
M_B^2\right)}{\mathcal{L}_{0}\left(s_0, M_B^2\right)}}. \label{mass}
\end{eqnarray}

The continuum threshold $s_0$ and Borel mass $M_B$ are the two most 
important parameters in a QCD sum rule analysis. The stability of QCD
sum rules requires a suitable working region. To obtain the Borel
window, one should study the OPE convergence and the pole
contribution. The criterion of OPE convergence determines a
lower bound on $M_B^2$ while the constraint of the pole contribution
leads to its upper bound. The pole contribution (PC) is defined as
\begin{eqnarray}
\text{PC}(s_0, M_B^2)=\frac{\mathcal{L}_{0}\left(s_0,
M_B^2\right)}{\mathcal{L}_{0}\left(\infty, M_B^2\right)}, \label{PC}
\end{eqnarray}
which is a function of $s_0$ and $M_B$. This definition comes from
the sum rules established in Eq.~\eqref{sumrule} and indicates the
contribution of the lowest lying resonance to the correlation
function. For the continuum threshold $s_0$, an optimized choice is
the value minimizing the variation of the extracted hadron mass
$m_X$ with the Borel mass $M_B^2$.

 \subsection{Input parameters}
So far, we have introduced the sum rules in Eq.~\eqref{sumrule} and
extracted the mass of the lowest lying ground hadron state in 
Eq.~\eqref{mass}. They are expressed as functions of the quark masses,
strong coupling $\alpha_s$, various QCD condensates, continuum 
threshold $s_0$ and Borel mass $M_B$. To perform a numerical analysis,
we adopt the following values of these parameters:
\cite{1985-Reinders-p1-1,2012-Beringer-p10001-10001,
2012-Narison-p259-263,2010-Narison-p559-559,2007-Kuhn-p192-215,2009-Chetyrkin-p74010-74010}:
\begin{eqnarray}
\nonumber && m_q=m_u=m_d=0, \non &&
m_s(2\,\text{GeV})=(101^{+29}_{-21})\text{ MeV} \, , \non
&&m_c(\mu=m_c)=\overline m_c=(1.23\pm 0.09)~\mbox{GeV}   \, , \non
&&m_b(\mu=m_b)=\overline m_b=(4.20\pm 0.07)~\mbox{GeV}   \, , \non
&&\qq=-(0.23\pm0.03)^3\text{ GeV}^3 \, , \non &&\qGqb=-M_0^2\qq\, ,
\non &&M_0^2=(0.8\pm0.2)\text{ GeV}^2 \, , \non &&\langle\bar
ss\rangle/\qq=0.8\pm0.2 \, , \non &&\GGb=(0.48 - 0.94)\text{GeV}^4\,
, \non && \GGGb=-(8.2\pm1.0)~\mbox{GeV}^2\langle\alpha_s GG\rangle,
\\ &&\JJb=-\frac{4}{3}g_s^4\qq^2,
\label{parameters}
\end{eqnarray}
in which we keep $m_u=m_d=0$ in the chiral limit. The charm and
bottom quark masses are the running masses in the $\overline{\rm
MS}$ scheme. One should be very cautious about the extra minus sign
in the values $\qGqb$ and $\GGGb$, which comes from our convention 
for the covariant derivative and strong coupling constant $g_s$.

For the conventional meson sum rules, the quark condensate $\qq$ and gluon condensate 
$\GGb$ are the dominant nonperturbative contributions to the correlation
functions and the contributions of the higher dimension condensates
are usually suppressed. However, the situations are different for
the exotic hadrons, such as the molecular and tetraquark states. In
these systems, the quark-gluon mixed condensate $\qGqb$, four-quark
condensate $\langle \bar qq\bar qq\rangle$, and even the dimension-8
condensate $\langle\bar qq\bar qg_s\sigma\cdot Gq\rangle$ will play
very important roles. 
The values of the four-quark condensate and
the dimension-8 condensate are estimated using the vacuum factorization
assumption \cite{1985-Bagan-p555-555,1984-Generalis-p85-85}
\begin{eqnarray}
\langle \bar qq\bar qq\rangle \sim \qq^2, \langle\bar qq\bar
qg_s\sigma\cdot Gq\rangle \sim \qq\qGqb.
\end{eqnarray}

For the hybrid charmonium and bottomonium analyses, the correlation
functions and the spectral densities are evaluated up to dimension
six tri-gluon condensate $\GGGb$ at the leading order of $\alpha_s$. 
The tri-gluon condensate $\GGGb$ was found to stabilize 
the mass sum rules in these systems \cite{2012-Qiao-p15005-15005,
2012-Harnett-p125003-125003,2012-Berg-p34002-34002,
2013-Chen-p19-19,2014-Chen-p25003-25003}. 
The strong coupling is then determined at the scale $\mu$ appropriate to the system by the evolution from the
$\tau$ and $Z$ masses, respectively:
\begin{eqnarray}
\alpha_s(\mu)&=&\frac{\alpha_s(M_{\tau})}{1+\frac{25\alpha_s(M_{\tau})}{12\pi}\log(\frac{\mu^2}{M_{\tau}^2})},
\quad \alpha_s(M_{\tau})=0.33; \label{alpha_cc}
\\ \alpha_s(\mu)&=&\frac{\alpha_s(M_{Z})}{1+\frac{23\alpha_s(M_{Z})}{12\pi}\log(\frac{\mu^2}{M_{Z}^2})}, \quad \alpha_s(M_{Z})=0.118,
\label{alpha_bb}
\end{eqnarray}
in which the $\tau$ and $Z$ masses, $\alpha_s(M_{\tau})$ and
$\alpha_s(M_{Z})$ are from the Particle Data
Group~\cite{2012-Beringer-p10001-10001}.

 \section{Construction of the interpolating currents}\label{sec:currents}
In general, there are two types of constructions of the four-quark
interpolating currents: tetraquark-type $(qq)(\bar q\bar q)$ and
molecular-type $(\bar qq)(\bar qq)$. However, these two
constructions are equivalent by using Fierz transformations 
\cite{2008-Chen-p54017-54017,2009-Jiao-p114034-114034}. In this
section, we construct all heavy tetraquark-type $(qQ)(\bar q\bar Q)$
and $(QQ)(\bar q\bar q)$ interpolating currents. These currents
have definite 
quantum numbers, flavor and color
structures. We will also introduce the quarkonium hybrid
interpolating currents with various quantum numbers.

 \subsection{Hidden charm/bottom tetraquark $Qq\bar Q\bar q$ interpolating currents}
To construct the diquark-antidiquark type of tetraquark currents, we
consider the six diquark fields and the corresponding antidiquark
fields as introduced in Table \ref{table1} and compose a six-order
matrix $O$. The elements of $O$ are the tetraquark operators, which 
are composed by multiplying a diquark and an antidiquark pair. 
The spins and
parities of the matrix elements with $J\le 1$ are listed in
Table~\ref{table2}.
\begin{center}
\renewcommand{\arraystretch}{1.3}
\begin{tabular*}{13cm}{cc|ccccccc}
\hlinewd{.8pt} Operators                           &
&   $\bar q_a\gamma_5C\bar Q_b^T$
                                    & $\bar q_aC\bar Q_b^T$             &   $\bar q_a\gamma_{\mu}\gamma_5C\bar Q_b^T$
                                    & $\bar q_a\gamma_{\mu}C\bar Q_b^T$ &   $\bar q_a\sigma_{\mu\nu}C\bar Q_b^T$
                                    & $\bar q_a\sigma_{\mu\nu}\gamma_5C\bar Q_b^T$ \\
\hline
                                    & $J^P$ &   $0^+$     &    $0^-$    &    $1^-$    &    $1^+$    &    $1^-$    &    $1^+$ \\
\hline
$q_a^TC\gamma_5Q_b$                 & $0^+$ &    $0^+$    &    $0^-$    &    $1^-$    &    $1^+$    &    $ - $    &    $ - $ \\
$q_a^TCQ_b$                         & $0^-$ &    $0^-$    &    $0^+$    &    $1^+$    &    $1^-$    &    $ - $    &    $ - $ \\
$q_a^TC\gamma_{\mu}\gamma_5Q_b$     & $1^-$ &    $1^-$    &    $1^+$    &    $0^+$    &    $0^-$    &    $1^+$    &    $1^-$ \\
$q_a^TC\gamma_{\mu}Q_b$             & $1^+$ &    $1^+$    &    $1^-$    &    $0^-$    &    $0^+$    &    $1^-$    &    $1^+$ \\
$q_a^TC\sigma_{\mu\nu}Q_b$          & $1^-$ &    $ - $    &    $ - $    &    $1^+$    &    $1^-$    &    $0^+$    &    $0^-$ \\
$q_a^TC\sigma_{\mu\nu}\gamma_5Q_b$  & $1^+$ &    $ - $    &    $ - $    &    $1^-$    &    $1^+$    &    $ - $    &    $ - $ \\
\hlinewd{.8pt}
\end{tabular*}
\tabcaption{The spins and parities of the elements of the matrix
$O$.} \label{table2}
\end{center}

For this operator matrix $O$, its charge-conjugation partner equals
to the transpose matrix $O^T$,
\begin{eqnarray}
\mathbb{C}O_{ij}\mathbb{C}^{-1}=O_{ji}. \label{matrix relation}
\end{eqnarray}

One notes that the operator $O_{66}$ is equivalent to $O_{55}$ while
$O_{65}$ is equivalent to $O_{56}$, including their spins and
parities. Thus we do not use $O_{66}$ and $O_{65}$ in the
construction of the currents. To compose a color singlet tetraquark
current, the color structure of the tetraquark is either $\mathbf 6
\otimes \mathbf {\bar 6}$ or $\mathbf { \bar 3 }\otimes \mathbf 3$,
which is denoted by $\mathbf  6 $ and $\mathbf 3 $ respectively.
With the relation in Eq.~\eqref{matrix relation}, we can define the
symmetric matrix $S$ and antisymmetric matrix $A$,
\begin{eqnarray}
S_{\mathbf 6}&=&O_{\mathbf 6}+O_{\mathbf 6}^T, S_{\mathbf 3}=O_{\mathbf 3}+O_{\mathbf 3}^T,\\
A_{\mathbf 6}&=&O_{\mathbf 6}-O_{\mathbf 6}^T, A_{\mathbf
3}=O_{\mathbf 3}-O_{\mathbf 3}^T,
\end{eqnarray}
in which $S_{\mathbf 6}, A_{\mathbf 6}$ have symmetric color
structures and $S_{\mathbf 3}, A_{\mathbf 3}$ have antisymmetric
color structures. It is easy to check that the tetraquark elements
of $S$ have even C-parities and the elements of $A$ have odd C-parities.
$A_{ii}=0$ indicates that the $J^{PC}=0^{+-}$ tetraquark currents
without derivatives do not exist \cite{2009-Jiao-p114034-114034}.
However, one can construct the tetraquark currents with such quantum
numbers by using a covariant derivative \cite{2013-Du-p33104-33104}.

Finally, we can obtain the tetraquark interpolating currents with
$J^{PC}=0^{-+}, 0^{--}, 1^{-+}, 1^{--}, 1^{++}$ and $1^{+-}$ from
the matrices $S_{\mathbf 6}, A_{\mathbf 6}$, $S_{\mathbf 3}$ and
$A_{\mathbf 3}$:
\begin{itemize}
\item The interpolating currents with $J^{PC}=0^{-+}$ and $0^{--}$ are:
\begin{eqnarray}
\nonumber J_1 &=&S_{21}^{\mathbf 6}(A_{21}^{\mathbf
6})=q_{a}^TCQ_{b}(\bar{q}_{a}\gamma_5C\bar{Q}^T_{b}+\bar{q}_{b}\gamma_5C\bar{Q}^T_{a})\pm
q_{a}^TC\gamma_5Q_{b}(\bar{q}_{a}C\bar{Q}^T_{b}+\bar{q}_{b}C\bar{Q}^T_{a})
\, , \non J_2 &=&O_{56}^{\mathbf
3}=q_{a}^TC\sigma_{\mu\nu}Q_{b}(\bar{q}_{a}\sigma^{\mu\nu}\gamma_5C\bar{Q}^T_{b}-\bar{q}_{b}\sigma^{\mu\nu}
\gamma_5C\bar{Q}^T_{a}) \, , \non J_3 &=&S_{43}^{\mathbf
6}(A_{43}^{\mathbf
6})=q_{a}^TC\gamma_{\mu}Q_{b}(\bar{q}_{a}\gamma^{\mu}\gamma_5C\bar{Q}^T_{b}+\bar{q}_{b}\gamma^{\mu}
\gamma_5C\bar{Q}^T_{a})\pm
q_{a}^TC\gamma_{\mu}\gamma_5Q_{b}(\bar{q}_{a}\gamma^{\mu}C\bar{Q}^T_{b}+\bar{q}_{b}
\gamma^{\mu}C\bar{Q}^T_{a}) \, , \non J_4 &=&S_{43}^{\mathbf
3}(A_{43}^{\mathbf
3})=q_{a}^TC\gamma_{\mu}Q_{b}(\bar{q}_{a}\gamma^{\mu}\gamma_5C\bar{Q}^T_{b}-\bar{q}_{b}\gamma^{\mu}
\gamma_5C\bar{Q}^T_{a})\pm
q_{a}^TC\gamma_{\mu}\gamma_5Q_{b}(\bar{q}_{a}\gamma^{\mu}C\bar{Q}^T_{b}-\bar{q}_{b}
\gamma^{\mu}C\bar{Q}^T_{a}) \, , \non J_5 &=&S_{21}^{\mathbf
3}(A_{21}^{\mathbf
3})=q_{a}^TCQ_{b}(\bar{q}_{a}\gamma_5C\bar{Q}^T_{b}-\bar{q}_{b}\gamma_5C\bar{Q}^T_{a})
\pm
q_{a}^TC\gamma_5Q_{b}(\bar{q}_{a}C\bar{Q}^T_{b}-\bar{q}_{b}C\bar{Q}^T_{a})
\, ,
\\
J_6 &=&O_{56}^{\mathbf
6}=q_{a}^TC\sigma_{\mu\nu}Q_{b}(\bar{q}_{a}\sigma^{\mu\nu}\gamma_5C\bar{Q}^T_{b}+\bar{q}_{b}\sigma^{\mu\nu}
\gamma_5C\bar{Q}^T_{a}) \, , \label{currents1}
\end{eqnarray}
where ``$S$'' and ``$+$'' in $J_1, J_3, J_4, J_5$ correspond to
$J^{PC}=0^{-+}$, ``$A$'' and ``$-$'' correspond to $J^{PC}=0^{--}$.
$J_2, J_6$ couple to the states with $J^{PC}=0^{-+}$.

\item The interpolating currents with $J^{PC}=1^{-+}$ and $1^{--}$ are:
\begin{eqnarray}
\nonumber J_{1\mu}&=&S_{13}^{\mathbf 6}(A_{13}^{\mathbf 6})
=q_{a}^TC\gamma_5Q_{b}(\bar{q}_{a}\gamma_{\mu}\gamma_5C\bar{Q}^T_{b}+\bar{q}_{b}\gamma_{\mu}\gamma_5C\bar{Q}^T_{a})
\pm
q_{a}^TC\gamma_{\mu}\gamma_5Q_{b}(\bar{q}_{a}\gamma_5C\bar{Q}^T_{b}+\bar{q}_{b}\gamma_5C\bar{Q}^T_{a})\,
, \non J_{2\mu}&=&S_{45}^{\mathbf 3}(A_{45}^{\mathbf 3})
=q_{a}^TC\gamma^{\nu}Q_{b}(\bar{q}_{a}\sigma_{\mu\nu}C\bar{Q}^T_{b}-\bar{q}_{b}\sigma_{\mu\nu}C\bar{Q}^T_{a})
\pm
q_{a}^TC\sigma_{\mu\nu}Q_{b}(\bar{q}_{a}\gamma^{\nu}C\bar{Q}^T_{b}-\bar{q}_{b}\gamma^{\nu}C\bar{Q}^T_{a})\,
, \non J_{3\mu}&=&S_{13}^{\mathbf 3}(A_{13}^{\mathbf 3})
=q_{a}^TC\gamma_5Q_{b}(\bar{q}_{a}\gamma_{\mu}\gamma_5C\bar{Q}^T_{b}-\bar{q}_{b}\gamma_{\mu}\gamma_5C\bar{Q}^T_{a})
\pm
q_{a}^TC\gamma_{\mu}\gamma_5Q_{b}(\bar{q}_{a}\gamma_5C\bar{Q}^T_{b}-\bar{q}_{b}\gamma_5C\bar{Q}^T_{a})\,
, \non J_{4\mu}&=&S_{45}^{\mathbf 6}(A_{45}^{\mathbf 6})
=q_{a}^TC\gamma^{\nu}Q_{b}(\bar{q}_{a}\sigma_{\mu\nu}C\bar{Q}^T_{b}+\bar{q}_{b}\sigma_{\mu\nu}C\bar{Q}^T_{a})
\pm
q_{a}^TC\sigma_{\mu\nu}Q_{b}(\bar{q}_{a}\gamma^{\nu}C\bar{Q}^T_{b}+\bar{q}_{b}\gamma^{\nu}C\bar{Q}^T_{a})\,
, \non J_{5\mu}&=&S_{24}^{\mathbf 6}(A_{24}^{\mathbf 6})
=q_{a}^TCQ_{b}(\bar{q}_{a}\gamma_{\mu}C\bar{Q}^T_{b}+\bar{q}_{b}\gamma_{\mu}C\bar{Q}^T_{a})
\pm
q_{a}^TC\gamma_{\mu}Q_{b}(\bar{q}_{a}C\bar{Q}^T_{b}+\bar{q}_{b}C\bar{Q}^T_{a})\,
, \non J_{6\mu}&=&S_{36}^{\mathbf 6}(A_{36}^{\mathbf 6})
=q_{a}^TC\gamma^{\nu}\gamma_5Q_{b}(\bar{q}_{a}\sigma_{\mu\nu}\gamma_5C\bar{Q}^T_{b}+\bar{q}_{b}\sigma_{\mu\nu}\gamma_5C\bar{Q}^T_{a})
\pm
q_{a}^TC\sigma_{\mu\nu}\gamma_5Q_{b}(\bar{q}_{a}\gamma^{\nu}\gamma_5C\bar{Q}^T_{b}+\bar{q}_{b}\gamma^{\nu}
\gamma_5C\bar{Q}^T_{a})\, , \non J_{7\mu}&=&S_{24}^{\mathbf
3}(A_{24}^{\mathbf 3})
=q_{a}^TCQ_{b}(\bar{q}_{a}\gamma_{\mu}C\bar{Q}^T_{b}-\bar{q}_{b}\gamma_{\mu}C\bar{Q}^T_{a})
\pm
q_{a}^TC\gamma_{\mu}Q_{b}(\bar{q}_{a}C\bar{Q}^T_{b}-\bar{q}_{b}C\bar{Q}^T_{a})\,
, \\ J_{8\mu}&=&S_{36}^{\mathbf 3}(A_{36}^{\mathbf 3})
=q_{a}^TC\gamma^{\nu}\gamma_5Q_{b}(\bar{q}_{a}\sigma_{\mu\nu}\gamma_5C\bar{Q}^T_{b}-\bar{q}_{b}\sigma_{\mu\nu}\gamma_5C\bar{Q}^T_{a})
\pm
q_{a}^TC\sigma_{\mu\nu}\gamma_5Q_{b}(\bar{q}_{a}\gamma^{\nu}\gamma_5C\bar{Q}^T_{b}-\bar{q}_{b}\gamma^{\nu}
\gamma_5C\bar{Q}^T_{a})\, , \label{currents2}
\end{eqnarray}
where ``$S$'' and ``$+$'' correspond to $J^{PC}=1^{-+}$, ``$A$'' and
``$-$'' correspond to $J^{PC}=1^{--}$.

\item The interpolating currents with $J^{PC}=1^{++}$ and $1^{+-}$
are:
\begin{eqnarray}
\nonumber J_{1\mu}&=&S_{23}^{\mathbf 6}(A_{23}^{\mathbf 6})
=q^T_aCQ_b(\bar{q}_a\gamma_{\mu}\gamma_5C\bar{Q}^T_b+\bar{q}_b\gamma_{\mu}\gamma_5C\bar{Q}^T_a)
\pm
q^T_aC\gamma_{\mu}\gamma_5Q_b(\bar{q}_aC\bar{Q}^T_b+\bar{q}_bC\bar{Q}^T_a)\,
, \non J_{2\mu}&=&S_{23}^{\mathbf 3}(A_{23}^{\mathbf 3})
=q^T_aCQ_b(\bar{q}_a\gamma_{\mu}\gamma_5C\bar{Q}^T_b-\bar{q}_b\gamma_{\mu}\gamma_5C\bar{Q}^T_a)
\pm
q^T_aC\gamma_{\mu}\gamma_5Q_b(\bar{q}_aC\bar{Q}^T_b-\bar{q}_bC\bar{Q}^T_a)\,
, \non J_{3\mu}&=&S_{14}^{\mathbf 6}(A_{14}^{\mathbf 6})
=q^T_aC\gamma_5Q_b(\bar{q}_a\gamma_{\mu}C\bar{Q}^T_b+\bar{q}_b\gamma_{\mu}C\bar{Q}^T_a)
\pm
q^T_aC\gamma_{\mu}Q_b(\bar{q}_a\gamma_5C\bar{Q}^T_b+\bar{q}_b\gamma_5C\bar{Q}^T_a)\,
, \non J_{4\mu}&=&S_{14}^{\mathbf 3}(A_{14}^{\mathbf 3})
=q^T_aC\gamma_5Q_b(\bar{q}_a\gamma_{\mu}C\bar{Q}^T_b-\bar{q}_b\gamma_{\mu}C\bar{Q}^T_a)
\pm
q^T_aC\gamma_{\mu}Q_b(\bar{q}_a\gamma_5C\bar{Q}^T_b-\bar{q}_b\gamma_5C\bar{Q}^T_a)\,
, \non J_{5\mu}&=&S_{46}^{\mathbf 6}(A_{46}^{\mathbf 6})
=q^T_aC\gamma^{\nu}Q_b(\bar{q}_a\sigma_{\mu\nu}\gamma_5C\bar{Q}^T_b+\bar{q}_b\sigma_{\mu\nu}\gamma_5C\bar{Q}^T_a)
\pm
q^T_aC\sigma_{\mu\nu}\gamma_5Q_b(\bar{q}_a\gamma^{\nu}C\bar{Q}^T_b+\bar{q}_b\gamma^{\nu}C\bar{Q}^T_a)\,
, \non J_{6\mu}&=&S_{46}^{\mathbf 3}(A_{46}^{\mathbf 3})
=q^T_aC\gamma^{\nu}Q_b(\bar{q}_a\sigma_{\mu\nu}\gamma_5C\bar{Q}^T_b-\bar{q}_b\sigma_{\mu\nu}\gamma_5C\bar{Q}^T_a)
\pm
q^T_aC\sigma_{\mu\nu}\gamma_5Q_b(\bar{q}_a\gamma^{\nu}C\bar{Q}^T_b-\bar{q}_b\gamma^{\nu}C\bar{Q}^T_a)\,
, \non J_{7\mu}&=&S_{35}^{\mathbf 6}(A_{35}^{\mathbf 6})
=q^T_aC\gamma^{\nu}\gamma_5Q_b(\bar{q}_a\sigma_{\mu\nu}C\bar{Q}^T_b+\bar{q}_b\sigma_{\mu\nu}C\bar{Q}^T_a)
\pm
q^T_aC\sigma_{\mu\nu}Q_b(\bar{q}_a\gamma^{\nu}\gamma_5C\bar{Q}^T_b+\bar{q}_b\gamma^{\nu}\gamma_5C\bar{Q}^T_a)\,
, \\ J_{8\mu}&=&S_{35}^{\mathbf 3}(A_{35}^{\mathbf 3})
=q^T_aC\gamma^{\nu}\gamma_5Q_b(\bar{q}_a\sigma_{\mu\nu}C\bar{Q}^T_b-\bar{q}_b\sigma_{\mu\nu}C\bar{Q}^T_a)
\pm
q^T_aC\sigma_{\mu\nu}Q_b(\bar{q}_a\gamma^{\nu}\gamma_5C\bar{Q}^T_b-\bar{q}_b
\gamma^{\nu}\gamma_5C\bar{Q}^T_a)\,, \label{currents3}
\end{eqnarray}
where ``$S$'' and ``$+$'' correspond to $J^{PC}=1^{++}$, ``$A$'' and
``$-$'' correspond to $J^{PC}=1^{+-}$.
\end{itemize}

All the currents in Eqs.~\eqref{currents1}--\eqref{currents3} should
contain both the $uc\bar u \bar c$ and $dc\bar d \bar c$ parts to
have the definite isospin and $G$-parity. However, we do not
differentiate the up and down quarks in our analysis due to the
$SU(2)$ flavor symmetry and denote them by $q$. Using these
currents, we calculate the correlation functions and the spectral
densities up to dimension eight. One can find the results of the
spectral densities in Refs.~\cite{2010-Chen-p105018-105018,2011-Chen-p34010-34010}. From these
expressions, the nonperturbative terms include the quark condensate
$\qq$, quark-gluon mixed condensate $\qGqb$, gluon condensate
$\GGb$, four-quark condensate $\qq^2$ and dimension-8 condensate
$\qq\qGqb$. The tri-gluon condensate $\GGGb$ is heavily suppressed
in the heavy tetraquark systems and can be neglected
\cite{2010-Chen-p105018-105018}. To study the $sQ\bar s\bar Q$
systems, we keep the $m_s$ dependent terms in the spectral densities.

 \subsection{Doubly charmed/bottomed tetraquark $QQ\bar q\bar q$ interpolating currents}
For the $QQ\bar q\bar q$ systems, the heavy quark pair $QQ$ has the
symmetric flavor structure $\mathbf {6_f}$. According to Table
\ref{table1}, its color structure is determined at the same time. To
compose a color singlet tetraquark operator, the heavy quark pair
should be multiplied by a light antiquark pair $\bar q\bar q$ with
the same color structure. Considering the Pauli principle, we can
obtain the following $QQ\bar q\bar q$ currents with $J^P=0^-, 0^+,
1^-$ and $1^+$:
\begin{itemize}
\item For the currents with $J^P=0^-$,
\begin{equation}
\begin{split}
\eta_1&=Q^T_aCQ_b(\bar{q}_a\gamma_5C\bar{q}_b^T+\bar{q}_b\gamma_5C\bar{q}^T_a),\\
\eta_2&=Q^T_aC\gamma_5Q_b(\bar{q}_aC\bar{q}_b^T+\bar{q}_bC\bar{q}^T_a),\\
\eta_3&=Q^T_aC\sigma_{\mu\nu}Q_b(\bar{q}_a
\sigma^{\mu\nu}\gamma_5C\bar{q}^T_b-\bar{q}_b\sigma^{\mu\nu}\gamma_5C\bar{q}^T_a),\\
\eta_4&=Q^T_aC\gamma_\mu Q_b(\bar{q}_a\gamma^\mu\gamma_5
C\bar{q}^T_b-
\bar{q}_b\gamma^\mu\gamma_5 C\bar{q}^T_a),\\
\eta_5&=Q^T_aC\gamma_\mu\gamma_5Q_b(\bar{q}_a\gamma^\mu
C\bar{q}^T_b+\bar{q}_b\gamma^\mu C\bar{q}^T_a),\label{current4}
\end{split}
\end{equation}
in which $\eta_1, \eta_2, \eta_3$ are isovector currents with
$[\mathbf{\bar 6_f}]_{\bar{q}\bar q}$ and $I=1$, $\eta_4, \eta_5$
are isoscalar currents with $[\mathbf{3_f}]_{\bar{q}\bar q}$ and
$I=0$.
\item For the currents with $J^P=0^+$,
\begin{equation}
\begin{split}
\eta_1&=Q^T_aCQ_b(\bar{q}_aC\bar{q}_b^T+\bar{q}_bC\bar{q}^T_b),\\
\eta_2&=Q^T_aC\gamma_5Q_b(\bar{q}_a\gamma_5C\bar{q}_b^T+\bar{q}_b\gamma_5C\bar{q}^T_b),\\
\eta_3&=Q^T_aC\gamma_\mu Q_b(\bar{q}_a\gamma^\mu C\bar{q}_b^T-\bar{q}_b \gamma^\mu C\bar{q}^T_b),\\
\eta_4&=Q^T_aC\gamma_\mu\gamma_5Q_b(\bar{q}_a\gamma^\mu \gamma_5C\bar{q}_b^T+\bar{q}_b\gamma^\mu \gamma_5C\bar{q}^T_b),\\
\eta_5&=Q^T_aC\sigma_{\mu\nu}Q_b(\bar{q}_a\sigma^{\mu\nu}C\bar{q}^T_b-\bar{q}_b\sigma^{\mu\nu}C\bar{q}^T_a),\label{current5}
\end{split}
\end{equation}
and all the scalar interpolating currents are isovector currents
with $[\mathbf{\bar 6_f}]_{\bar{q}\bar q}$ and $I=1$.
\item For the currents with $J^P=1^-$,
\begin{equation}
\begin{split}
\eta_1&=Q^T_aC\gamma_\mu\gamma_5 Q_b(\bar{q}_a\gamma_5C\bar{q}_b^T+\bar{q}_b\gamma_5C\bar{q}^T_a),\\
\eta_2&=Q^T_aC\gamma_5Q_b(\bar{q}_a\gamma_\mu\gamma_5 C\bar{q}_b^T+\bar{q}_b\gamma_\mu\gamma_5 C\bar{q}_a^T),\\
\eta_3&=Q^T_aC\sigma_{\mu\nu}Q_b(\bar{q}_a\gamma^\nu C\bar{q}^T_b-\bar{q}_b\gamma^\nu C\bar{q}^T_a),\\
\eta_4&=Q^T_aC\gamma^\nu Q_b(\bar{q}_a\sigma_{\mu\nu}
C\bar{q}^T_b-\bar{q}_b\sigma_{\mu\nu}C\bar{q}^T_a),\\
\eta_5&=Q^T_aC\gamma_\mu Q_b(\bar{q}_aC\bar{q}_b^T-\bar{q}_bC\bar{q}^T_a),\\
\eta_6&=Q^T_aCQ_b(\bar{q}_a\gamma_\mu C\bar{q}_b^T+\bar{q}_b\gamma_\mu C\bar{q}_a^T),\\
\eta_7&=Q^T_aC\sigma_{\mu\nu}\gamma_5Q_b(\bar{q}_a\gamma^\nu\gamma_5 C\bar{q}^T_b-\bar{q}_b\gamma^\nu\gamma_5 C\bar{q}^T_a),\\
\eta_8&=Q^T_aC\gamma^\nu\gamma_5
Q_b(\bar{q}_a\sigma_{\mu\nu}\gamma_5
C\bar{q}^T_b+\bar{q}_b\sigma_{\mu\nu}\gamma_5C\bar{q}^T_a),\label{current6}
\end{split}
\end{equation}
in which $\eta_1, \eta_2, \eta_3, \eta_4$ are isovector currents
with $[\mathbf{\bar 6_f}]_{\bar{q}\bar q}$ and $I=1$, $\eta_5,
\eta_6, \eta_7, \eta_8$ are isoscalar currents with
$[\mathbf{3_f}]_{\bar{q}\bar q}$ and $I=0$.
\item For the currents with $J^P=1^+$,
\begin{equation}
\begin{split}
\eta_1&=Q^T_aC\gamma_\mu\gamma_5 Q_b(\bar{q}_aC\bar{q}_b^T+\bar{q}_bC\bar{q}^T_a),\\
\eta_2&=Q^T_aCQ_b(\bar{q}_a\gamma_\mu\gamma_5 C\bar{q}_b^T+\bar{q}_b\gamma_\mu\gamma_5 C\bar{q}_a^T),\\
\eta_3&=Q^T_aC\sigma_{\mu\nu}\gamma_5 Q_b(\bar{q}_a\gamma^\nu C\bar{q}^T_b-\bar{q}_b\gamma^\nu C\bar{q}^T_a),\\
\eta_4&=Q^T_aC\gamma^\nu Q_b(\bar{q}_a\sigma_{\mu\nu}\gamma_5
C\bar{q}^T_b-\bar{q}_b\sigma_{\mu\nu}\gamma_5
C\bar{q}^T_a),\\
\eta_5&=Q^T_aC\gamma_\mu Q_b(\bar{q}_a\gamma_5C\bar{q}_b^T-\bar{q}_b\gamma_5C\bar{q}^T_a),\\
\eta_6&=Q^T_aC\gamma_5Q_b(\bar{q}_a\gamma_\mu C\bar{q}_b^T+\bar{q}_b\gamma_\mu C\bar{q}_a^T),\\
\eta_7&=Q^T_aC\sigma_{\mu\nu} Q_b(\bar{q}_a\gamma^\nu\gamma_5 C\bar{q}^T_b-\bar{q}_b\gamma^\nu\gamma_5 C\bar{q}^T_a),\\
\eta_8&=Q^T_aC\gamma^\nu\gamma_5
Q_b(\bar{q}_a\sigma_{\mu\nu}C\bar{q}^T_b+\bar{q}_b\sigma_{\mu\nu}
C\bar{q}^T_a), \label{current7}
\end{split}
\end{equation}
in which $\eta_1, \eta_2, \eta_3, \eta_4$ are isovector currents
with $[\mathbf{\bar 6_f}]_{\bar{q}\bar q}$ and $I=1$, $\eta_5,
\eta_6, \eta_7, \eta_8$ are isoscalar currents with
$[\mathbf{3_f}]_{\bar{q}\bar q}$ and $I=0$.
\end{itemize}
In fact, the two pieces in the parenthesis of 
Eqs.~\eqref{current4}--\eqref{current7} are equivalent to each other. We
keep both of them here to illustrate the color symmetry explicitly.
For these currents, we show their quark contents, spins, parities,
isospins and the flavor symmetries of the light quark pair in Table
\ref{table3}. The results of the spectral densities for these doubly
charmed/bottomed tetraquark currents can be found in Ref.~\cite{2013-Du-p14003-14003}.

\begin{center}
\renewcommand{\arraystretch}{1.3}
\begin{tabular*}{13cm}{c|c|c|c|c|c|c}
\hlinewd{.8pt}
Quark Content     & ~$[\bar{q}\bar q]_\mathbf{f}$  ~ &  ~~  I  ~~   &   $J^P=0^-$   &   $J^P=0^+$   &  $J^P=1^-$   &  $J^P=1^+$ \\
\hline $QQ\bar q\bar q$  & $\mathbf{\bar 6_f}$ &    1        &
$\eta_1,\eta_2,\eta_3$  &  $\eta_1,\eta_2,\eta_3,\eta_4,\eta_5$
                                 &  $\eta_1,\eta_2,\eta_3,\eta_4$  &  $\eta_1,\eta_2,\eta_3,\eta_4$ \\
$QQ\bar s\bar s$  & $\mathbf{\bar 6_f}$ &    0        &
$\eta_1,\eta_2,\eta_3$  &  $\eta_1,\eta_2,\eta_3,\eta_4,\eta_5$
                                 &  $\eta_1,\eta_2,\eta_3,\eta_4$  &  $\eta_1,\eta_2,\eta_3,\eta_4$ \\
$QQ\bar q\bar s$  & $\mathbf{\bar 6_f}$ &  $1/2$      &
$\eta_1,\eta_2,\eta_3,$
                                 &  $\eta_1,\eta_2,\eta_3,\eta_4,\eta_5$
                                 &  $\eta_1,\eta_2,\eta_3,\eta_4,$ &  $\eta_1,\eta_2,\eta_3,\eta_4,$\\
                  & $\mathbf{3_f}$      &&  $\eta_4,\eta_5$&&$\eta_5,\eta_6,\eta_7,\eta_8$&$\eta_5,\eta_6,\eta_7,\eta_8$ \\
$QQ\bar u\bar d$  & $\mathbf{3_f}$      &    0        &
$\eta_4,\eta_5$          &  $-$
                                 &  $\eta_5,\eta_6,\eta_7,\eta_8$  &  $\eta_5,\eta_6,\eta_7,\eta_8$ \\
\hlinewd{.8pt}
\end{tabular*}
\tabcaption{Properties of the doubly charmed/bottomed tetraquark
currents.} \label{table3}
\end{center}

 \subsection{Open-flavor tetraquark $bc\bar{q}\bar{q}$ and $qc\bar{q}\bar{b}$ interpolating currents}
In Ref.~\cite{2013-Chen-p-b}, the open-flavor tetraquark $bc\bar{q}\bar{q}$ and $qc\bar{q}\bar{b}$ interpolating currents
with quantum numbers $J^P=0^+, 1^+$ were constructed as follows. 
\begin{itemize}
\item For the scalar $bc\bar q\bar q$ system with $J^P=0^+$:
\begin{equation}
\begin{split}
J_1&=b^T_aC\gamma_5c_b(\bar{q}_a\gamma_5C\bar{q}^T_b+\bar{q}_b\gamma_5C\bar{q}^T_a),\\
J_2&=b^T_aC\gamma_\mu c_b(\bar{q}_a\gamma^\mu
C\bar{q}^T_b+\bar{q}_b\gamma^\mu C\bar{q}^T_a),
\\
J_3&=b^T_aC\gamma_5c_b(\bar{q}_a\gamma_5C\bar{q}^T_b-\bar{q}_b\gamma_5C\bar{q}^T_a),\\
J_4&=b^T_aC\gamma_\mu c_b(\bar{q}_a\gamma^\mu
C\bar{q}^T_b-\bar{q}_b\gamma^\mu C\bar{q}^T_a), \label{currentbc1}
\end{split}
\end{equation}
where $J_1, J_2$ have the symmetric color structure
$[\mathbf{6_c}]_{bc} \otimes [\mathbf{ \bar 6_c}]_{\bar{q}\bar q}$
and $J_3, J_4$ have the antisymmetric color structure $[\mathbf{\bar
3_c}]_{bc} \otimes [\mathbf{3_c}]_{\bar{q}\bar q}$.
\end{itemize}

\begin{itemize}
\item For the vector $bc\bar q\bar q$ system with $J^P=1^+$:
\begin{equation}
\begin{split}
J_{1\mu}&=b^T_aC\gamma_5c_b(\bar{q}_a\gamma_\mu C\bar{q}_b^T+\bar{q}_b\gamma_\mu C\bar{q}_a^T),\\
J_{2\mu}&=b^T_aC\gamma_\mu c_b(\bar{q}_a\gamma_5C\bar{q}_b^T+\bar{q}_b\gamma_5C\bar{q}^T_a),\\
J_{3\mu}&=b^T_aC\gamma_5c_b(\bar{q}_a\gamma_\mu C\bar{q}_b^T-\bar{q}_b\gamma_\mu C\bar{q}_a^T),\\
J_{4\mu}&=b^T_aC\gamma_\mu
c_b(\bar{q}_a\gamma_5C\bar{q}_b^T-\bar{q}_b\gamma_5C\bar{q}^T_a),
\label{currentbc2}
\end{split}
\end{equation}
where $J_{1\mu}, J_{2\mu}$ have the symmetric color structure
$[\mathbf{6_c}]_{bc} \otimes [\mathbf{ \bar 6_c}]_{\bar{q}\bar q}$
and $J_{3\mu}, J_{4\mu}$ have the antisymmetric color structure $[\mathbf{\bar
3_c}]_{bc} \otimes [\mathbf{3_c}]_{\bar{q}\bar q}$.
\end{itemize}

\begin{itemize}
\item For the scalar $cq\bar b\bar q$ system with $J^P=0^+$:
\begin{equation}
\begin{split}
J_1&=q^T_aC\gamma_5c_b(\bar{q}_a\gamma_5C\bar{b}^T_b+\bar{q}_b\gamma_5C\bar{b}^T_a),\\
J_2&=q^T_aC\gamma_\mu c_b(\bar{q}_a\gamma^\mu
C\bar{b}^T_b+\bar{q}_b\gamma^\mu C\bar{b}^T_a),
\\
J_3&=q^T_aC\gamma_5c_b(\bar{q}_a\gamma_5C\bar{b}^T_b-\bar{q}_b\gamma_5C\bar{b}^T_a),\\
J_4&=q^T_aC\gamma_\mu c_b(\bar{q}_a\gamma^\mu
C\bar{b}^T_b-\bar{q}_b\gamma^\mu C\bar{b}^T_a), \label{currentbc3}
\end{split}
\end{equation}
where $J_1, J_2$ have the symmetric color structure
$[\mathbf{6_c}]_{qc} \otimes [\mathbf{ \bar 6_c}]_{\bar{q}\bar b}$
and $J_3, J_4$ have the antisymmetric color structure $[\mathbf{\bar
3_c}]_{qc} \otimes [\mathbf{3_c}]_{\bar{q}\bar b}$. 
\end{itemize}

\begin{itemize}
\item For the vector $cq\bar b\bar q$ system with $J^P=1^+$:
\begin{equation}
\begin{split}
J_{1\mu}&=q^T_aC\gamma_5c_b(\bar{q}_a\gamma_\mu C\bar{b}_b^T+\bar{q}_b\gamma_\mu C\bar{b}_a^T),\\
J_{2\mu}&=q^T_aC\gamma_\mu c_b(\bar{q}_a\gamma_5C\bar{b}_b^T+\bar{q}_b\gamma_5C\bar{b}^T_a),\\
J_{3\mu}&=q^T_aC\gamma_5c_b(\bar{q}_a\gamma_\mu C\bar{b}_b^T-\bar{q}_b\gamma_\mu C\bar{b}_a^T),\\
J_{4\mu}&=q^T_aC\gamma_\mu
c_b(\bar{q}_a\gamma_5C\bar{b}_b^T-\bar{q}_b\gamma_5C\bar{b}^T_a),
\label{currentbc4}
\end{split}
\end{equation}
where $J_{1\mu}, J_{2\mu}$ have the symmetric color structure
$[\mathbf{6_c}]_{qc} \otimes [\mathbf{ \bar 6_c}]_{\bar{q}\bar b}$
and $J_{3\mu}, J_{4\mu}$ have the antisymmetric color structure $[\mathbf{\bar
3_c}]_{qc} \otimes [\mathbf{3_c}]_{\bar{q}\bar b}$.
\end{itemize}

The $cs\bar b\bar s$ tetraquark currents with $J^P=0^+, 1^+$ are the same as the 
$cq\bar b\bar q$ currents in Eqs.~\eqref{currentbc3} and \eqref{currentbc4} respectively, 
by replacing the light quark $q$ by the strange quark $s$. For the $bc\bar s\bar s$ 
systems, the flavor structure of $\bar s\bar s$ pair is symmetric. Thus the color structures 
for the diquark fields $s_a^TC\gamma_5s_b$ and $s_a^TC\gamma_{\mu}s_b$ are symmetric 
$\mathbf{6_c}$ and antisymmetric $\mathbf{\bar 3_c}$, respectively. As a result, only
$J_1$, $J_4$ in Eq.~\eqref{currentbc1} and $J_{2\mu}$, $J_{3\mu}$ in
Eq.~\eqref{currentbc2} survive in the $bc\bar s\bar s$ system. The spectral densities 
for these open-flavor tetraquark currents were calculated and listed in 
Ref.~\cite{2013-Chen-p-b}.

 \subsection{Quarkonium hybrid $\bar QGQ$ interpolating currents}
The quarkonium hybrids were originally studied in 
Refs.~\cite{1985-Govaerts-p215-215,1985-Govaerts-p575-575,1987-Govaerts-p674-674}
in the QCD sum rules method. However, it was shown that only the
hybrid channels with $J^{PC}=0^{--}, 0^{++}, 1^{+-}, 1^{++}, 2^{++}$
gave stable mass sum rules while the sum rules in the
$J^{PC}=0^{-+}, 0^{+-}, 1^{-+}, 1^{--}, 2^{-+}$ channels were
unstable. Recently, new efforts on the quarkonium
hybrid mesons have found that the dimension six
tri-gluon condensate can stabilize the hybrid sum rules
\cite{2012-Qiao-p15005-15005, 2012-Harnett-p125003-125003,
2012-Berg-p34002-34002,2013-Chen-p19-19,2014-Chen-p25003-25003}.

To study the hybrid correlation functions, we consider the following
interpolating currents with various quantum numbers:
\begin{eqnarray}
\nonumber
J^{(1)}_{\mu}&=&g_s\bar Q_1\frac{\lambda^a}{2}\gamma^{\nu}G^a_{\mu\nu}Q_2,~~~~~~~J^{P(C)}=1^{-(+)}, 0^{+(+)},
\\
J^{(2)}_{\mu}&=&g_s\bar Q_1\frac{\lambda^a}{2}\gamma^{\nu}\gamma_5G^a_{\mu\nu}Q_2,~~~~J^{P(C)}=1^{+(-)}, 0^{-(-)}, \label{currents}
\non
J^{(3)}_{\mu\nu}&=&g_s\bar Q_1\frac{\lambda^a}{2}\sigma_{\mu}^{\alpha}\gamma_5 G^a_{\alpha\nu}Q_2,~~~~J^{P(C)}=2^{-(+)}, 1^{+(+)}, 1^{-(+)}, 0^{-(+)}\,,
\end{eqnarray}
in which $Q_1$ and $Q_2$ are the heavy quark fields with masses $m_1$ and $m_2$, $g_s$ is the strong coupling constant, $\lambda^a$ 
are the Gell-Mann SU(3) matrices and $G^a_{\mu\nu}$ is the gluon field strength. It should be understood that the operators in Eq.~\eqref{currents} 
contain the hidden charm/bottom hybrid currents with $Q_1=Q_2$ carrying C-parities in the parentheses and the open-flavor $\bar bGc$ 
hybrid currents with $Q_1\neq Q_2$ carrying no definite C-parities. By replacing $G^a_{\mu\nu}$ with 
$\tilde G^a_{\mu\nu}=\frac{1}{2}\epsilon_{\mu\nu\alpha\beta}G^{\alpha\beta,a}$, one can also obtain the operators 
$\tilde J^{(1)}_{\mu}, \tilde J^{(2)}_{\mu}, \tilde J^{(3)}_{\mu\nu}$ coupling to the hybrid states with the opposite parities to 
$J^{(1)}_{\mu}, J^{(2)}_{\mu}, J^{(3)}_{\mu\nu}$ respectively. 
The correlation functions and spectral densities are calculated up to dimension six tri-gluon condensate at leading order in $\alpha_s$. 
We will discuss the importance of the tri-gluon condensate in the next section. The results of the spectral densities are listed in 
Refs.~\cite{2013-Chen-p19-19,2014-Chen-p25003-25003}.

 \section{Mass spectrum of the quarkoniumlike tetraquark $qQ\bar q\bar Q$ states}\label{sec:quarkoniumlike}
 The QCD sum rule study of the $qQ\bar q\bar Q$ systems was performed
 in Refs.~\cite{2010-Chen-p105018-105018,2011-Chen-p34010-34010}, using the
 interpolating currents in Eqs.~\eqref{currents1}--\eqref{currents3}. As introduced
 in Sec.~\ref{sec:intro}, the charmoniumlike tetraquark states are good candidates for 
 some newly observed $XYZ$ resonances. For example, the $Y(4660)$ meson was first
 observed by Belle in the $e^+e^-\to \gamma_{ISR}Y(4660)\to \gamma_{ISR}\psi(2S)\pi^+\pi^-$ process
 \cite{2007-Wang-p142002-142002}. Since it was observed in the initial state radiation (ISR) process,
 its quantum number is $J^{PC}=1^{--}$.
 The tetraquark interpolating current for such quantum numbers is
 given in Eq.~\eqref{currents2}.

\begin{center}
\begin{tabular}{c}
\scalebox{0.9}{\includegraphics{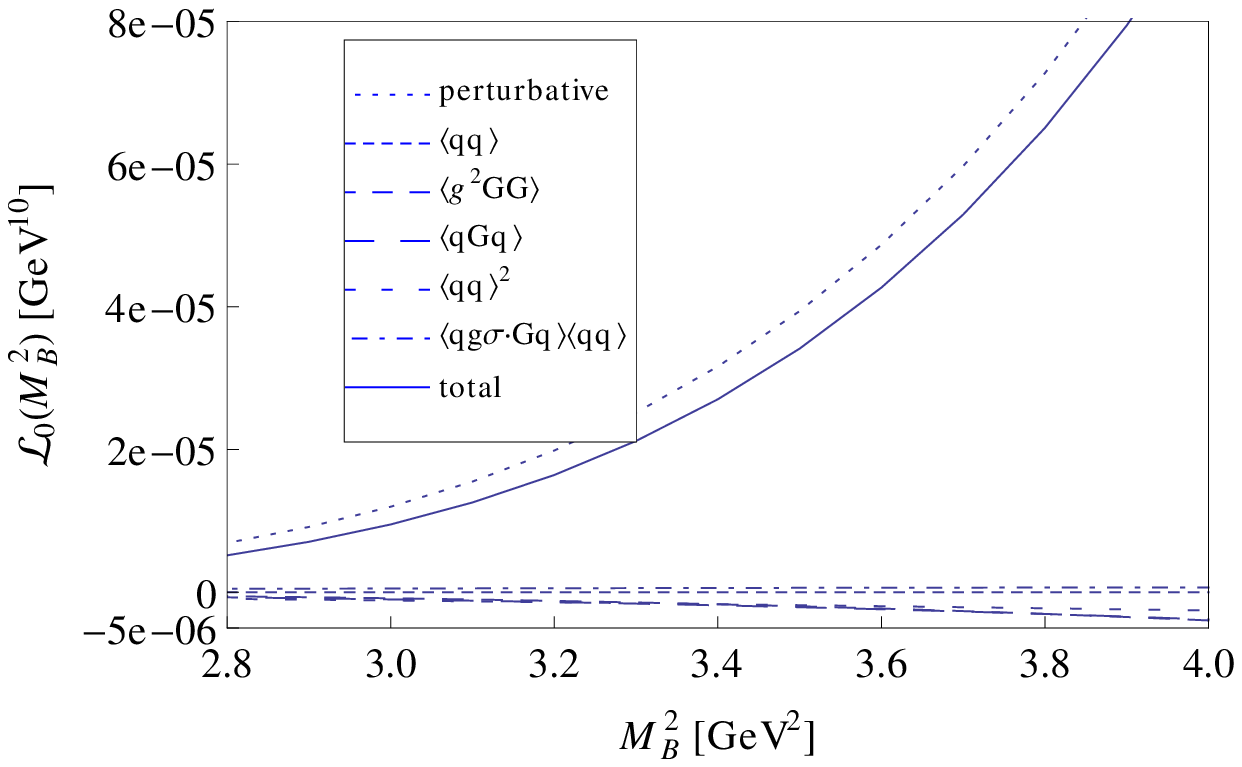}}
\end{tabular}
\figcaption{OPE convergence for $J_{1\mu}$ with $J^{PC}=1^{--}$
in the $qc\bar q\bar c$ systems.} \label{Y(4660)OPE}
\end{center}

For the $qc\bar q\bar c$ currents $J_{1\mu}$ with $J^{PC}=1^{--}$ in Eq.~\eqref{currents2},
 we show the convergence of the OPE
 series in Fig.~\ref{Y(4660)OPE}. We see that the four-quark condensate $\qq^2$ is the
 most important nonperturbative contribution
 to the correlation function. In fact, the quark condensate $\qq$ is proportional to
 the light quark mass and vanishes in the 
 limit $m_q=0$. 
 The OPE convergence is very good in the region $M_B^2\geq 2.9$ GeV$^2$.
 To study the pole contribution defined
 in Eq.~\eqref{PC}, one should determine the continuum threshold $s_0$. In the left 
 portion  of Fig.~\ref{Y(4660)mass}, the hadron mass
 $m_X$ is extracted as a function of $s_0$. One can find a plateau
 in the region $10$ GeV$^2$ $\leq s_0 \leq 14$ GeV$^2$. However, this is a non-physical 
 artifact because the spectral density of the sum rules defined
 in Eq.~\eqref{sumrule} is negative 
 in this region. The variation of $m_X$ with the
 Borel mass $M_B^2$ is weak around $s_0=25$ GeV$^2$. Using this value
 of the continuum threshold, we study the pole
 contribution and require that PC be larger than $40\%$, which results in
 an upper bound of the Borel mass $M_{\mbox{max}}^2=3.6$ GeV$^2$.

\begin{center}
\begin{tabular}{lr}
\scalebox{0.7}{\includegraphics{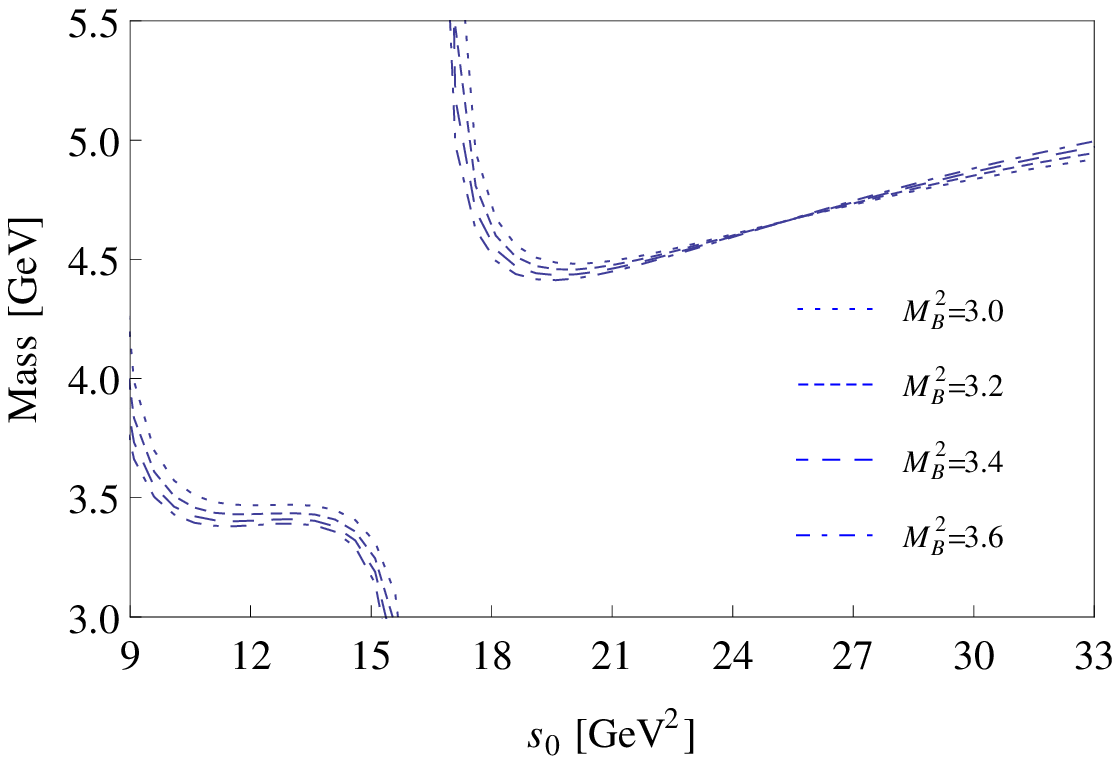}}&
\scalebox{0.7}{\includegraphics{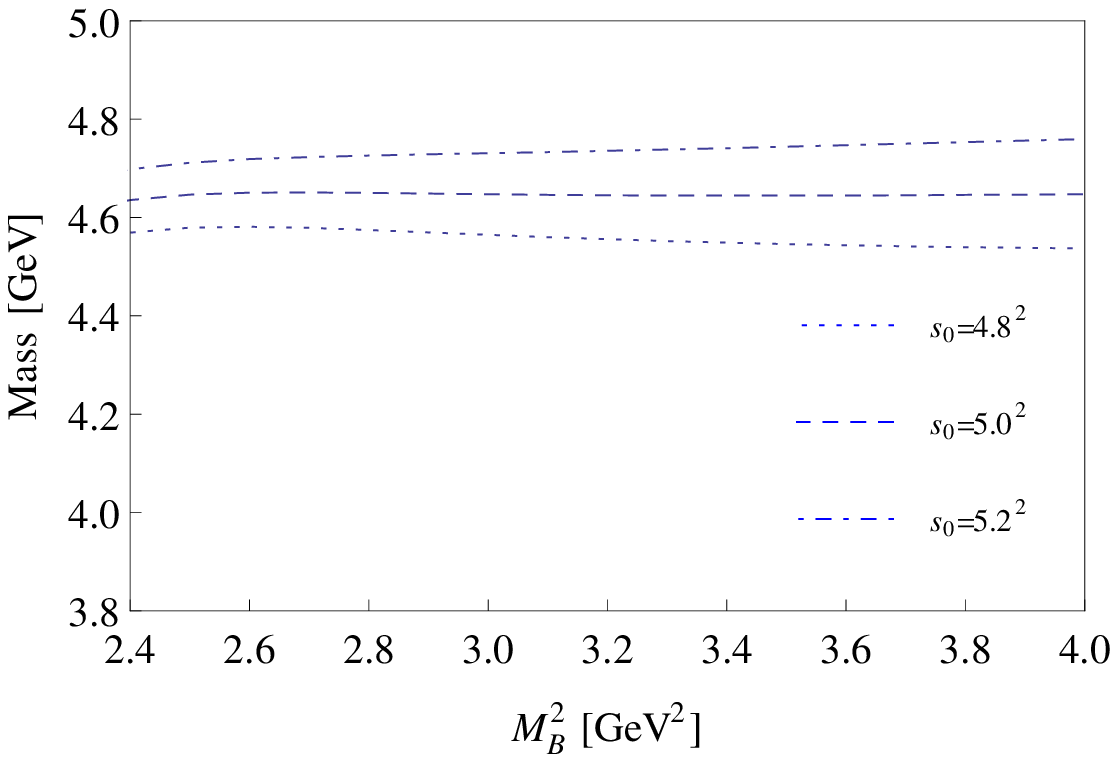}}
\end{tabular}
\figcaption{Variations of $m_X$ with $s_0$ and $M_B^2$ for
$J_{1\mu}$ with $J^{PC}=1^{--}$ in the $qc\bar q\bar c$ systems.}
\label{Y(4660)mass}
\end{center}

In the right portion of Fig.~\ref{Y(4660)mass}, we show the variation
of $m_X$ with $M_B^2$ for $J_{1\mu}$ with $J^{PC}=1^{--}$ in the
$qc\bar q\bar c$ systems. The curves are very stable in the Borel
window $2.9$ GeV$^2$ $\leq M_B^2 \leq 3.6$ GeV$^2$. The hadron mass
is then extracted as
\begin{eqnarray}
m_X=(4.64\pm 0.09) \mbox{GeV},
\end{eqnarray}
which is consistent with the mass of the $Y(4660)$ meson. This
result implies a possible tetraquark interpretation of $Y(4660)$.
However, the tetraquark state can decay easily into two meson final
states via the rearrangement mechanism so it is very difficult to
explain the small decay width of the $Y(4660)$ meson. The $Y(4660)$ was
also considered as a $f_0(980)\psi^{\prime}$ bound
state \cite{2008-Guo-p26-29}, a baryonium
state \cite{2008-Qiao-p75008-75008}, a $5~ ^3S_1 c\bar c$
state \cite{2008-Ding-p14033-14033} and a $sc\bar s\bar c$ tetraquark
state \cite{2008-Ebert-p399-405,2009-Albuquerque-p53-66}.

We can study the corresponding $sc\bar s\bar c$ state by replacing
the light quarks in the interpolating current by a strange quark.
The quark condensate $\langle \bar ss\rangle$ is now proportional to
the strange quark mass $m_s$. Its contribution is larger than the
four-quark condensate $\langle \bar ss\rangle^2$ and becomes the
dominant power correction for the $1^{--}$ $sc\bar s\bar c$ system.
The extracted mass of this state is \cite{2011-Chen-p34010-34010}:
\begin{eqnarray}
m_{X^s}=(4.92\pm 0.10) \mbox{GeV},
\end{eqnarray}
which is $0.28$ GeV higher than the $1^{--}$ $qc\bar q\bar c$ state.

Properties of the bottomoniumlike analogues are very similar due to
the heavy quark symmetry. Replacing the charm quark by the bottom
quark in the current and repeating the same analysis procedures, we
can also extract the masses of the $qb\bar q\bar b$ and $sb\bar
s\bar b$ tetraquark states with $J^{PC}=1^{--}$. After performing
the QCD sum rule analysis for all the interpolating currents in
Eqs.~\eqref{currents1}--\eqref{currents3}, we obtained the numerical
results for $J^{PC}=0^{--}, 0^{-+}, 1^{-+}, 1^{--}, 1^{++}$ and
$1^{+-}$ tetraquark systems in Tables \ref{table0--},
\ref{table0-+}, \ref{table1-+}, \ref{table1--}, \ref{table1++} and
\ref{table1+-}, respectively \cite{2010-Chen-p105018-105018,2011-Chen-p34010-34010}. 
Only the errors
from the uncertainty of the threshold values and variation of the
Borel parameter are taken into account. Other possible error
sources include the truncation of the OPE series and the
uncertainty of the quark masses, condensate values and so on.
We only collect the numerical results
from the interpolating currents which lead to stable mass sum rules and
reliable mass predictions in these tables. 
The tetraquark states with $J^{PC}=0^{--}$ and $1^{-+}$ are exotic
states. They do not mix with the conventional charmonium and
bottomonium states because these quantum numbers are not accessible
to simple $q\bar q$ states.

\begin{center}
\renewcommand{\arraystretch}{1.3}
\begin{tabular*}{12cm}{cccccc}
\hlinewd{.8pt}
                   & Currents & $s_0(\mbox{GeV}^2)$ & $[M^2_{\mbox{min}}$, $M^2_{\mbox{max}}](\mbox{GeV}^2)$& $m_X$\mbox{(GeV)}      & PC(\%)\\
\hline
 $J^{PC}=0^{--}$                    & $J_2$      &  25                  & $2.4-3.7$                         & $4.55\pm0.11$    & 46.3  \\

                     & $J_4$      &  25                  & $2.4-3.7$                         & $4.55\pm0.11$    & 45.9 \\
\hlinewd{.8pt}
\end{tabular*}
\tabcaption{Numerical results for the $qc\bar q\bar c$ tetraquark
states with $J^{PC}=0^{--}$. \label{table0--}}
\end{center}
\begin{center}
\renewcommand{\arraystretch}{1.3}
\begin{tabular*}{12cm}{cccccc}
\hlinewd{.8pt}
                   & Currents & $s_0(\mbox{GeV}^2)$ & $[M^2_{\mbox{min}}$, $M^2_{\mbox{max}}](\mbox{GeV}^2)$& $m_X$\mbox{(GeV)}      & PC(\%)\\
\hline
                     & $J_2$      &  27                  & $2.4-4.1$                         & $4.72\pm0.10$    & 53.8 \\
$J^{PC}=0^{-+}$
                     & $J_5$      &  25                  & $2.4-3.7$                         & $4.55\pm0.11$    & 45.9 \\
                     & $J_{6}$   &  27                  & $2.4-4.2$                         & $4.67\pm0.10$    & 56.8 \\
\hlinewd{.8pt}
\end{tabular*}
\tabcaption{Numerical results for the $qc\bar q\bar c$ tetraquark
states with $J^{PC}=0^{-+}$. \label{table0-+}}
\end{center}

\begin{center}
\renewcommand{\arraystretch}{1.3}
\begin{tabular*}{12cm}{cccccc}
\hlinewd{.8pt}
                   & Currents & $s_0(\mbox{GeV}^2)$&$[M^2_{\mbox{min}}$,$M^2_{\mbox{max}}](\mbox{GeV}^2)$&$m_X$\mbox{(GeV)}&PC(\%)\\
\hline
                        & $J_{6\mu}$      &  $5.1^2$             & $2.9-3.9$              & $4.67\pm0.10$    & 50.2  \\
$qc\bar q\bar c$ system & $J_{7\mu}$      &  $5.2^2$             & $2.9-4.2$              & $4.77\pm0.10$    & 47.4  \\
                        & $J_{8\mu}$      &  $4.9^2$             & $2.9-3.4$              & $4.53\pm0.10$    & 46.3
\vspace{5pt}\\
                        & $J_{1\mu}$      &  $5.0^2$             & $2.9-3.4$              & $4.67\pm0.10$    & 44.3  \\
                        & $J_{2\mu}$      &  $5.0^2$             & $2.9-3.4$              & $4.65\pm0.09$    & 45.6  \\
                        & $J_{3\mu}$      &  $4.9^2$             & $2.9-3.3$              & $4.54\pm0.10$    & 44.4  \\
                        & $J_{4\mu}$      &  $5.1^2$             & $2.9-3.7$              & $4.72\pm0.09$    & 44.8  \\
$sc\bar s\bar c$ system & $J_{5\mu}$      &  $5.0^2$             & $2.9-3.6$              & $4.62\pm0.10$    & 42.8  \\
                        & $J_{6\mu}$      &  $5.3^2$             & $2.9-4.3$              & $4.84\pm0.10$    & 47.3  \\
                        & $J_{7\mu}$      &  $5.3^2$             & $2.9-4.3$              & $4.87\pm0.10$    & 46.2  \\
                        & $J_{8\mu}$      &  $5.2^2$             & $2.9-4.1$              & $4.77\pm0.10$    & 44.1
\vspace{5pt}\\
                        & $J_{6\mu}$      &  $11.0^2$            & $7.2-8.6$              & $10.53\pm0.11$   & 44.2  \\
$qb\bar q\bar b$ system & $J_{7\mu}$      &  $11.0^2$            & $7.2-8.6$              & $10.53\pm0.10$   & 44.1  \\
                        & $J_{8\mu}$      &  $11.0^2$            & $7.2-8.6$              & $10.49\pm0.11$   & 44.7
\vspace{5pt}\\
                        & $J_{4\mu}$      &  $11.0^2$            & $7.2-8.1$              & $10.62\pm0.10$   & 41.2  \\
                        & $J_{5\mu}$      &  $11.0^2$            & $7.2-8.4$              & $10.56\pm0.10$   & 43.8  \\
$sb\bar s\bar b$ system & $J_{6\mu}$      &  $11.0^2$            & $7.2-8.3$              & $10.63\pm0.10$   & 42.4  \\
                        & $J_{7\mu}$      &  $11.0^2$            & $7.2-8.3$              & $10.62\pm0.09$   & 42.5  \\
                        & $J_{8\mu}$      &  $11.0^2$            & $7.2-8.3$              & $10.59\pm0.10$   & 43.1  \\
\hlinewd{.8pt}
\end{tabular*}
\tabcaption{Numerical results for the $qc\bar q\bar c$, $sc\bar
s\bar c$, $qb\bar q\bar b$ and $sb\bar s\bar b$ tetraquark states
with $J^{PC}=1^{-+}$.\label{table1-+}}
\end{center}
\begin{center}
\renewcommand{\arraystretch}{1.3}
\begin{tabular*}{12cm}{cccccc}
\hlinewd{.8pt}
                   & Currents & $s_0(\mbox{GeV}^2)$&$[M^2_{\mbox{min}}$,$M^2_{\mbox{max}}](\mbox{GeV}^2)$&$m_X$\mbox{(GeV)}&PC(\%)\\
\hline
                        & $J_{1\mu}$      &  $5.0^2$         & $2.9-3.6$           & $4.64\pm0.09$     & 44.1  \\
$qc\bar q\bar c$ system & $J_{4\mu}$      &  $5.0^2$         & $2.9-3.6$           & $4.61\pm0.10$     & 46.4  \\
                        & $J_{7\mu}$      &  $5.2^2$         & $2.9-4.1$           & $4.74\pm0.10$     & 47.3
\vspace{5pt} \\
                        & $J_{1\mu}$      &  $5.4^2$         & $2.8-4.5$           & $4.92\pm0.10$     & 50.3  \\
                        & $J_{2\mu}$      &  $5.0^2$         & $2.8-3.5$           & $4.64\pm0.09$     & 48.6  \\
$sc\bar s\bar c$ system & $J_{3\mu}$      &  $4.9^2$         & $2.8-3.4$           & $4.52\pm0.10$     & 45.6  \\
                        & $J_{4\mu}$      &  $5.4^2$         & $2.8-4.5$           & $4.88\pm0.10$     & 51.7  \\
                        & $J_{7\mu}$      &  $5.3^2$         & $2.8-4.3$           & $4.86\pm0.10$     & 46.0  \\
                        & $J_{8\mu}$      &  $4.8^2$         & $2.8-3.1$           & $4.48\pm0.10$     & 43.2
\vspace{5pt} \\
$qb\bar q\bar b$ system & $J_{7\mu}$      &  $11.0^2$        &
$7.2-8.5$           & $10.51\pm0.10$    & 45.8
\vspace{5pt} \\
                        & $J_{1\mu}$      &  $11.0^2$        & $7.2-8.3$           & $10.60\pm0.10$    & 47.0  \\
                        & $J_{2\mu}$      &  $11.0^2$        & $7.2-8.4$           & $10.55\pm0.11$    & 43.6  \\
$sb\bar s\bar b$ system & $J_{3\mu}$      &  $11.0^2$        & $7.2-8.4$           & $10.55\pm0.10$    & 43.7  \\
                        & $J_{4\mu}$      &  $11.0^2$        & $7.2-8.4$           & $10.53\pm0.11$    & 44.3  \\
                        & $J_{7\mu}$      &  $11.0^2$        & $7.2-8.2$           & $10.62\pm0.10$    & 42.0  \\
                        & $J_{8\mu}$      &  $11.0^2$        & $7.2-8.4$           & $10.53\pm0.10$    & 44.1  \\
\hlinewd{.8pt}
\end{tabular*}
\tabcaption{Numerical results for the $qc\bar q\bar c$, $sc\bar
s\bar c$, $qb\bar q\bar b$ and $sb\bar s\bar b$ tetraquark states
with $J^{PC}=1^{--}$.\label{table1--}}
\end{center}
\begin{center}
\renewcommand{\arraystretch}{1.3}
\begin{tabular*}{12cm}{cccccc}
\hlinewd{.8pt}
                   & Currents & $s_0(\mbox{GeV}^2)$&$[M^2_{\mbox{min}}$,$M^2_{\mbox{max}}](\mbox{GeV}^2)$&$m_X$\mbox{(GeV)}&PC(\%)\\
\hline
$qc\bar q\bar c$ system & $J_{3\mu}$         &  $4.6^2$         & $3.0-3.4$           & $4.19\pm0.10$     & 47.3 \\
                        & $J_{4\mu}$         &  $4.5^2$         & $3.0-3.3$           & $4.03\pm0.11$     & 46.8
\vspace{5pt} \\
$sc\bar s\bar c$ system & $J_{3\mu}$         &  $4.6^2$         & $3.0-3.4$           & $4.22\pm0.10$     & 45.7  \\
                        & $J_{4\mu}$         &  $4.5^2$         & $3.0-3.3$           & $4.07\pm0.10$     & 44.4
\vspace{5pt} \\
                        & $J_{3\mu}$         &  $10.9^2$        & $8.5-9.5$           & $10.32\pm0.09$    & 47.0 \\
$qb\bar q\bar b$ system & $J_{4\mu}$         &  $10.8^2$        & $8.5-9.2$           & $10.22\pm0.11$    & 44.6 \\
                        & $J_{7\mu}$         &  $10.7^2$        & $7.8-8.4$           & $10.14\pm0.10$    & 44.8  \\
                        & $J_{8\mu}$         &  $10.7^2$        & $7.8-8.4$           & $10.14\pm0.09$    & 44.8
\vspace{5pt} \\
                        & $J_{3\mu}$         &  $10.9^2$        & $8.5-9.5$           & $10.34\pm0.09$    & 46.1 \\
$sb\bar s\bar b$ system & $J_{4\mu}$         &  $10.8^2$        & $8.5-9.1$           & $10.25\pm0.10$    & 43.3 \\
                        & $J_{7\mu}$         &  $10.8^2$        & $7.5-8.6$           & $10.24\pm0.11$    & 47.1  \\
                        & $J_{8\mu}$         &  $10.8^2$        & $7.5-8.6$           & $10.24\pm0.10$    & 47.1  \\
\hlinewd{.8pt}
\end{tabular*}
\tabcaption{Numerical results for the $qc\bar q\bar c$, $sc\bar
s\bar c$, $qb\bar q\bar b$ and $sb\bar s\bar b$ tetraquark states
with $J^{PC}=1^{++}$.\label{table1++}}
\end{center}
\begin{center}
\renewcommand{\arraystretch}{1.3}
\begin{tabular*}{12cm}{cccccc}
\hlinewd{.8pt}
                   & Currents & $s_0(\mbox{GeV}^2)$&$[M^2_{\mbox{min}}$,$M^2_{\mbox{max}}](\mbox{GeV}^2)$&$m_X$\mbox{(GeV)}&PC(\%)\\
\hline
                        & $J_{3\mu}$         &  $4.6^2$            & $3.0-3.4$           & $4.16\pm0.10$     & 46.2  \\
$qc\bar q\bar c$ system & $J_{4\mu}$         &  $4.5^2$            & $3.0-3.3$           & $4.02\pm0.09$     & 44.6  \\
                        & $J_{5\mu}$         &  $4.5^2$            & $3.0-3.4$           & $4.00\pm0.11$     & 46.0  \\
                        & $J_{6\mu}$         &  $4.6^2$            & $3.0-3.4$           & $4.14\pm0.09$     & 47.0
\vspace{5pt} \\
                        & $J_{3\mu}$         &  $4.7^2$            & $3.0-3.6$           & $4.24\pm0.10$     & 49.6  \\
$sc\bar s\bar c$ system & $J_{4\mu}$         &  $4.6^2$            & $3.0-3.5$           & $4.12\pm0.11$     & 47.3  \\
                        & $J_{5\mu}$         &  $4.5^2$            & $3.0-3.3$           & $4.03\pm0.11$     & 44.2  \\
                        & $J_{6\mu}$         &  $4.6^2$            & $3.0-3.4$           & $4.16\pm0.11$     & 46.0
\vspace{5pt} \\
                        & $J_{3\mu}$         &  $10.6^2$           & $7.5-8.5$           & $10.08\pm0.10$    & 45.9  \\
$qb\bar q\bar b$ system & $J_{4\mu}$         &  $10.6^2$           & $7.5-8.5$           & $10.07\pm0.10$    & 46.2  \\
                        & $J_{5\mu}$         &  $10.6^2$           & $7.5-8.4$           & $10.05\pm0.10$    & 45.3  \\
                        & $J_{6\mu}$         &  $10.7^2$           & $7.5-8.7$           & $10.15\pm0.10$    & 47.6
\vspace{5pt} \\
                        & $J_{3\mu}$         &  $10.6^2$           & $7.5-8.3$           & $10.11\pm0.10$    & 43.8  \\
$sb\bar s\bar b$ system & $J_{4\mu}$         &  $10.6^2$           & $7.5-8.4$           & $10.10\pm0.10$    & 44.1  \\
                        & $J_{5\mu}$         &  $10.6^2$           & $7.5-8.3$           & $10.08\pm0.10$    & 43.7  \\
                        & $J_{6\mu}$         &  $10.7^2$           & $7.5-8.5$           & $10.18\pm0.10$    & 46.5  \\
\hlinewd{.8pt}
\end{tabular*}
\tabcaption{Numerical results for the $qc\bar q\bar c$, $sc\bar
s\bar c$, $qb\bar q\bar b$ and $sb\bar s\bar b$ tetraquark states
with $J^{PC}=1^{+-}$.\label{table1+-}}
\end{center}

The $X(3872)$ is the first observed $XYZ$ state
\cite{2003-Choi-p262001-262001} and its quantum number was assigned
as $J^{PC}=1^{++}$\cite{2013-Aaij-p222001-222001}. The mass and
decay mode of $X(3872)$ are very different from that of the $2^3P_1$
$c\bar c$ state. To date, the possible interpretations of $X(3872)$
include the molecular state
\cite{2009-Liu-p411-428,2008-Liu-p63-73,2004-Swanson-p197-202,
2004-Swanson-p189-195, 2004-Close-p119-123,
2008-Thomas-p34007-34007, 2009-Fernandez-Carames-p222001-222001},
tetraquark state \cite{2007-Matheus-p14005-14005,
2007-Maiani-p182003-182003, 2006-Ebert-p214-219}, cusp
\cite{2004-Bugg-p8-14}, hybrid charmonium \cite{2003-Close-p210-216}
and mixed scenarios
\cite{2013-Chen-p45027-45027,2009-Matheus-p56002-56002}. In Table
\ref{table1++}, the masses of the $1^{++}$ $qc\bar q\bar c$ states
are $m_X=4.0 - 4.2$ GeV, in rough agreement with the mass of the
$X(3872)$, considering the uncertainties.

Recently, BESIII collaboration discovered a charged charmoniumlike
resonance $Z_c(4025)$ in the process $e^+e^-\to (D^*\bar
D^*)^{\pm}\pi^{\mp}$ \cite{2013-Ablikim-p-}. The quantum numbers of
$Z_c(4025)$ was $I^G(J^P)=1^+(1^+)$ because it was also observed in
$h_c\pi$ channel. Its neutral partner carries the negative
C-parity. Therefore its quantum number is $J^{PC}=1^{+-}$. The
extracted masses of the $qc\bar q\bar c$ states with $J^{PC}=1^{+-}$
in Table \ref{table1+-} and $1^{++}$ in Table \ref{table1++} are
about $m_X=4.0 - 4.2$ GeV, which supports $Z_c(4025)$ as a
$I^G(J^P)=1^+(1^+)$ tetraquark candidate.
The $Z_c(4025)$ were also studied as a $D^*\bar D^*$ 
molecular state \cite{2014-Chen-p2773-2773,2013-He-p2635-2635,2014-Wang-p2761-2761,2013-Cui-p-}, 
a $[cu][\bar c\bar d]$ tetraquark with the quantum numbers $J^P=2^+$ \cite{2013-Qiao-p-}.

 \section{Mass spectrum of the doubly charmed/bottomed tetraquark $QQ\bar q\bar q$ states}\label{sec:doubly}
The interpolating currents of the doubly charmed/bottomed $QQ\bar
q\bar q$ systems are listed in 
Eqs.~\eqref{current4}--\eqref{current7}. For all the isovector $QQ\bar
q\bar q$ and isoscalar $QQ\bar u\bar d$ systems, the dominant power
corrections are the four-quark condensate $\qq^2$. Both the quark
condensate and the quark-gluon mixed condensate are proportional to
the light quark mass and hence are chirally suppressed.

In Fig.~\ref{fig0-}, we show the variations of $m_X$ with $M_B$ and
$s_0$ for the current $\eta_3$ with $(I, J^P)=(1, 0^-)$ in
$cc\bar{q}\bar{q}$ system. The mass curves for the different values
of the Borel mass intersect at $s_0=23$ GeV$^2$, at which the
variation of $m_X$ with $M_B^2$ is very weak. After studying the OPE
convergence and the pole contribution, we obtain the Borel window
$2.6$ GeV$^2$ $\leq M_B^2 \leq 3.6$ GeV$^2$ in which the mass curves
are very stable, as shown in the left part of Fig.~\ref{fig0-}. The
mass was extracted around $m_X=(4.47\pm 0.12)$ GeV
\cite{2013-Du-p14003-14003}. 
The situation is very different from
the $(I, J^P)=(1, 0^-)$ in the $cc\bar{s}\bar{s}$ systems. The quark
condensate $\langle \bar ss\rangle$ becomes the dominant power
correction in these systems by keeping the $m_s$ dependent terms
in the spectral densities. To compare with the $cc\bar{q}\bar{q}$
system, we show the variations of $m_{X^s}$ with $M_B^2$ and $s_0$
for the current $\eta_3^s$ in Fig.~\ref{fig0-s}. The quark
condensate and the quark-gluon mixed condensate enhanced the pole
contribution, which resulted in a broader Borel window $2.7$ GeV$^2$
$\leq M_B^2 \leq 4.3$ GeV$^2$. The extracted mass was
$m_{X^s}=(4.79\pm 0.17)$ GeV, which is about $2m_s$ higher than the
$cc\bar{q}\bar{q}$ state \cite{2013-Du-p14003-14003}.

\begin{center}
\begin{tabular}{lr}
\scalebox{0.89}{\includegraphics{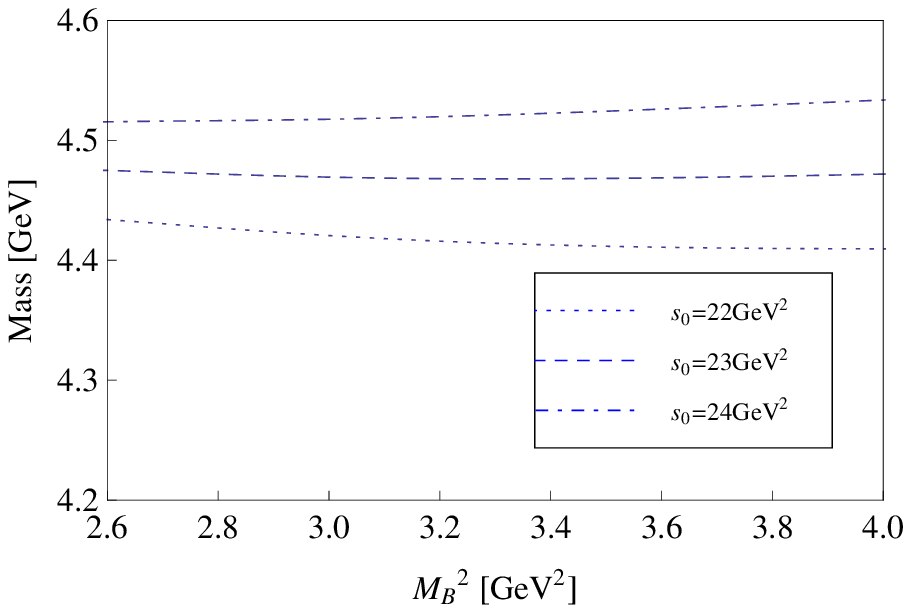}}&
\scalebox{0.89}{\includegraphics{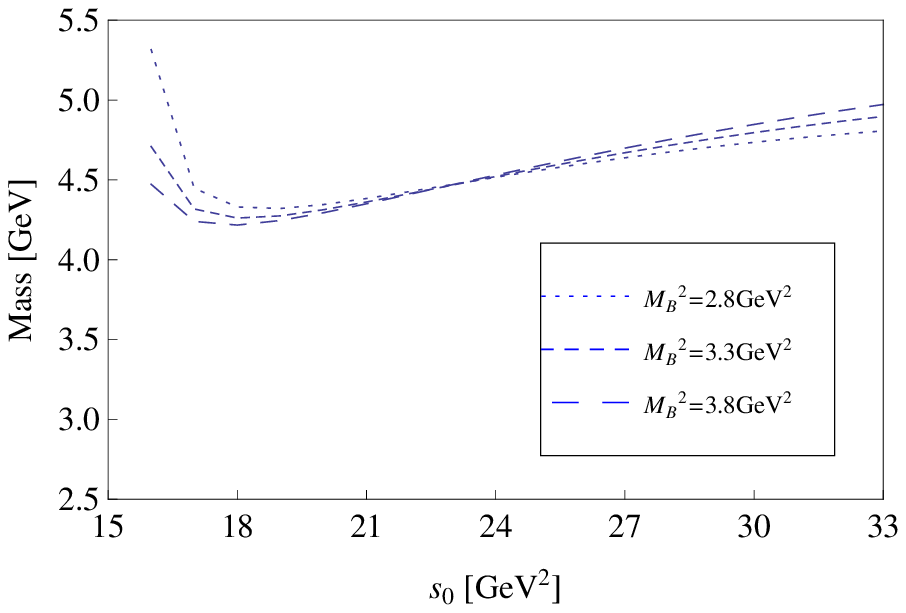}}
\end{tabular}
\figcaption{Variations of $m_X$ with $M_B$ and $s_0$ for the current
$\eta_3$ with $(I, J^P)=(1, 0^-)$ in $cc\bar{q}\bar{q}$ system.}
\label{fig0-}
\end{center}
\begin{center}
\begin{tabular}{lr}
\scalebox{0.7}{\includegraphics{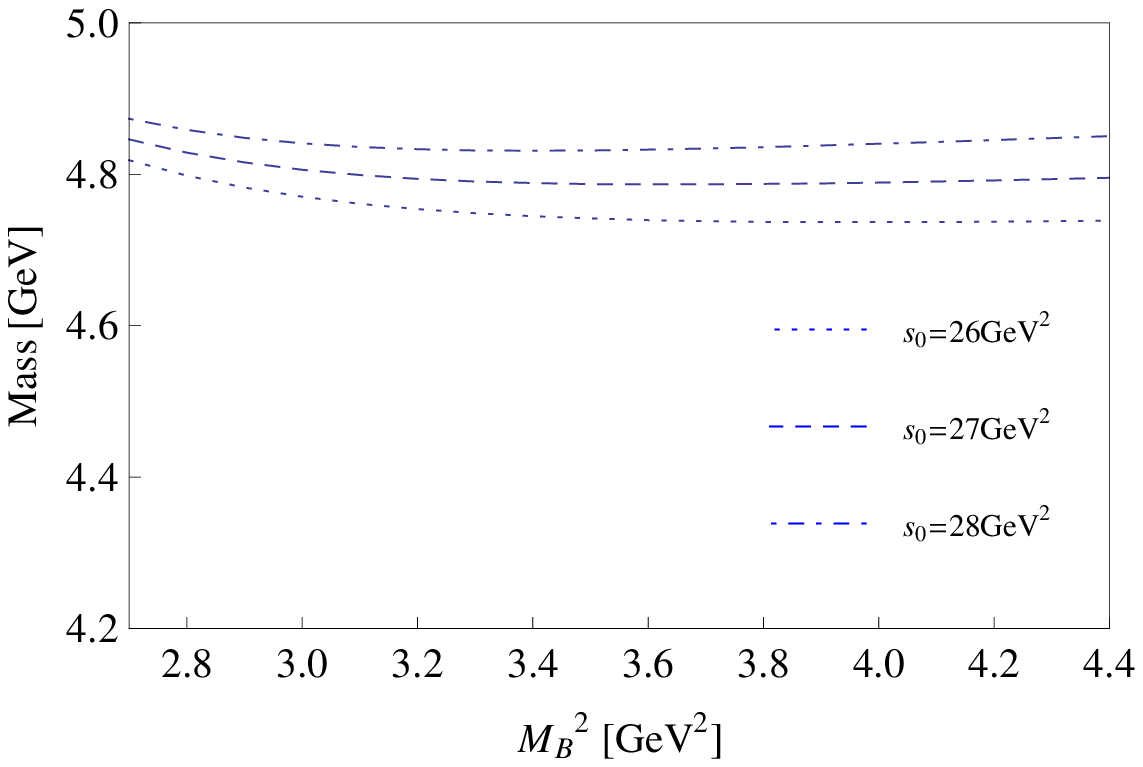}}&
\scalebox{0.7}{\includegraphics{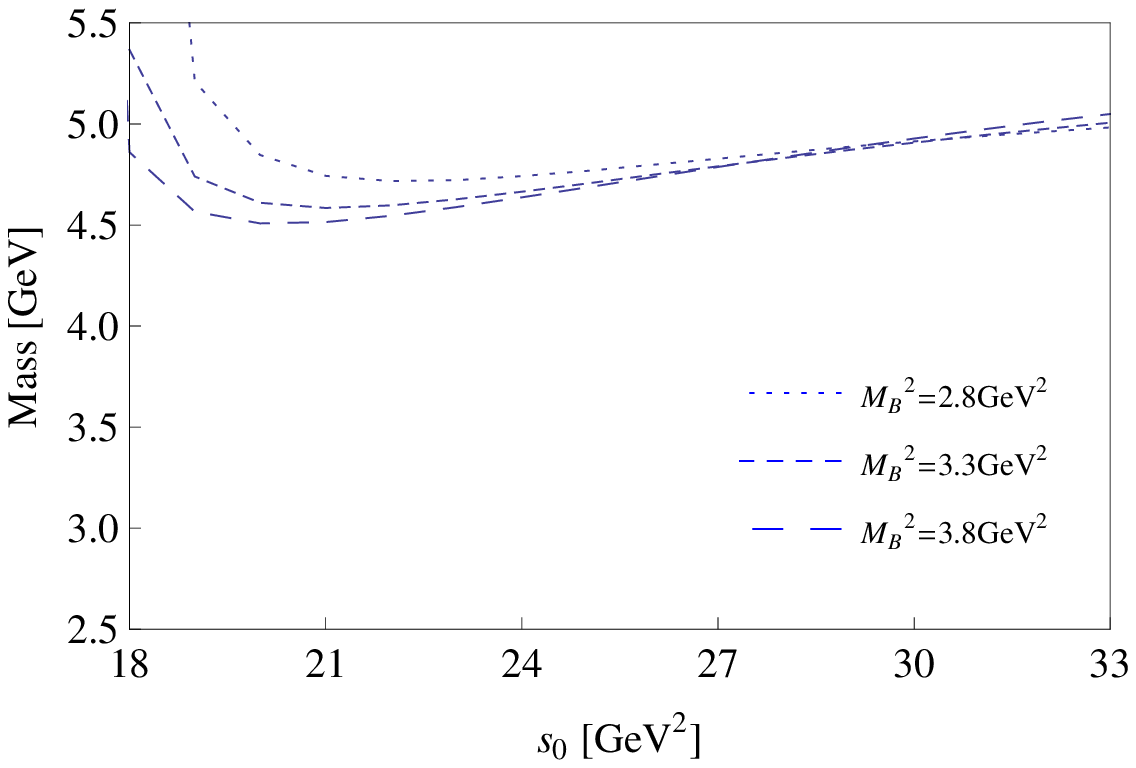}}
\end{tabular}
\figcaption{Variations of $m_{X^s}$ with $M_B$ and $s_0$ for the
current $\eta_3^s$ with $(I, J^P)=(0, 0^-)$ in $cc\bar{s}\bar{s}$
system.} \label{fig0-s}
\end{center}

\begin{center}
\renewcommand{\arraystretch}{1.2}
\begin{tabular*}{16cm}{lcccccccc}
\hlinewd{.8pt}
  &Current & $s_0$  & [$M^2_{Bmin},M^2_{Bmax}$] & $M^2_B$ & $m_X $ & ~~~~ PC~~~~ & ~~~~$f_X$~~~~ & open charm/bottom \\
                    &  & ($\mbox{GeV}^2$) & ($\mbox{GeV}^2$) &($\mbox{GeV}^2$)& $(\mbox{GeV})$ &(\%)&($\mbox{GeV}^5$) &threshold $(\mbox{GeV})$\\
                    \hline
$cc\bar{q}\bar{q}$  &$\eta_1$ & 24 & $3.0- 3.9$ & 3.4 & $4.43\pm0.12$  & 41.2  & 0.0674 & \\
                    &$\eta_3$ & 23 & $2.6- 3.6$ & 3.1 & $4.47\pm0.12$  & 42.6  & 0.312  & 3.872
                    \vspace{5pt} \\
$cc\bar{u}\bar{d}$  &$\eta_4$ & 22 & $2.7- 3.4$ & 3.0 & $4.43\pm0.13$  & 38.4  & 0.0870 & \\
                    &$\eta_5$ & 23 & $2.5- 3.7$ & 3.2 & $4.41\pm0.14$  & 41.5  & 0.106  &
                   \vspace{5pt} \\
                    &$\eta_1$ & 24 & $2.9- 3.8$ & 3.4 & $4.45\pm0.16$  & 40.1 & 0.0489 & \\
                    &$\eta_2$ & 24 & $2.7- 3.7$ & 3.4 & $4.68\pm0.12$  & 43.1 & 0.106 &
                    \vspace{5pt} \\
$cc\bar{q}\bar{s}$  &$\eta_3$ & 26 & $3.0- 4.4$ & 3.8 & $4.71\pm0.14$  & 40.6 & 0.245 & 3.975\\
                    &$\eta_4$ & 25 & $2.6- 4.1$ & 3.4 & $4.64\pm0.13$  & 44.9 & 0.136 & \\
                    &$\eta_5$ & 24 & $2.6- 4.1$ & 3.4 & $4.50\pm0.16$  & 45.9 & 0.124 &
                    \vspace{5pt} \\
$cc\bar{s}\bar{s}$  &$\eta_1$ & 25 & $2.8- 4.0$ & 3.4 & $4.46\pm0.13$  & 44.3 & 0.0731 & 4.081 \\
                    &$\eta_3$ & 27 & $2.7- 4.3$ & 3.4 & $4.79\pm0.17$  & 47.7 & 0.558  &
                    \vspace{5pt} \\
$bb\bar{q}\bar{q}$  &$\eta_1$ & 125 & $7.0- 9.6$ & 8.0 & $10.6\pm0.3$  & 48.6 & 0.207 &\\
                    &$\eta_3$ & 120 & $6.8- 9.4$ & 8.0 & $10.5\pm0.3$  & 43.7 & 1.60  & 10.60
                    \vspace{5pt} \\
$bb\bar{u}\bar{d}$  &$\eta_5$ & 115 & $7.0- 8.1$ & 7.5 &
$10.3\pm0.2$  & 36.5 & 0.367 &
                    \vspace{5pt} \\
                    &$\eta_1$ & 124 & $7.2- 9.6$ & 8.5 & $10.6\pm0.2$  & 40.8 & 0.188 &\\
$bb\bar{q}\bar{s}$  &$\eta_3$ & 120 & $7.2- 9.1$ & 8.0 & $10.6\pm0.2$  & 40.2 & 0.853 & 10.69\\
                    &$\eta_5$ & 115 & $7.0- 7.9$ & 7.5 & $10.4\pm0.2$  & 33.8 & 0.378 &
                    \vspace{5pt} \\
$bb\bar{s}\bar{s}$  &$\eta_1$ & 125 & $6.7- 9.6$ & 8.0 & $10.6\pm0.3$  & 47.9 & 0.286 &10.78\\
                    &$\eta_3$ & 120 & $6.6- 8.9$ & 8.0 & $10.6\pm0.3$  & 38.0 & 1.74  & \\
\hlinewd{.8pt}
\end{tabular*}
\tabcaption{The numerical results for the doubly-charmed/bottomed
$QQ\bar q\bar q$, $QQ\bar u\bar d$, $QQ\bar q\bar s$ and $QQ\bar
s\bar s$ systems with $J^P=0^-$.} \label{table0-}
\end{center}

\begin{center}
\renewcommand{\arraystretch}{1.3}
\begin{tabular*}{16cm}{lcccccccc}
\hlinewd{.8pt}
  &Current & $s_0$  & [$M^2_{Bmin},M^2_{Bmax}$] & $M^2_B$ & $m_X $ &~~~~ PC ~~~~& ~~~~$f_X$~~~~ & open charm/bottom \\
                    &  & ($\mbox{GeV}^2$) & ($\mbox{GeV}^2$) &($\mbox{GeV}^2$)& $(\mbox{GeV})$ &(\%)&($\mbox{GeV}^5$) &threshold $(\mbox{GeV})$\\
                    \hline
$cc\bar{q}\bar{s}$  &$\eta_2$ & 22 & $2.8- 3.6$ & 3.2 & $4.16\pm0.14$  & 39.0  & 0.0548 & 3.833\\
                    &$\eta_3$ & 20 & $2.6- 3.4$ & 3.0 & $4.02\pm0.18$   & 39.3 & 0.0561 &
                    \vspace{5pt} \\
$cc\bar{s}\bar{s}$  & $\eta_1$ & 28 & $3.2-4.1$ & 3.4 & $5.05\pm0.15$  & 43.3 & 0.136 & 3.937\\
                    &$\eta_2$ & 22 & $2.6-3.8$ & 3.2 & $4.27\pm0.11$ & 43.2  & 0.0933 &
                    \vspace{5pt} \\
                    & $\eta_2$ & 120 & $7.0- 9.8$ & 8.2 & $10.3\pm0.3$  & 48.2 & 0.590 & \\
$bb\bar{q}\bar{q}$  & $\eta_3$ & 115 & $6.9- 9.0$ & 8.0 & $10.2\pm0.3$  & 40.3 & 0.539 & 10.56\\
                    & $\eta_5$ & 115 & $6.7- 8.8$ & 8.0 & $10.2\pm0.3$  & 39.4& 1.10   &
                    \vspace{5pt} \\
                    & $\eta_3$ & 115 & $6.5- 8.8$ & 8.0 & $10.2\pm0.3$  & 40.3 & 0.398 & \\
$bb\bar{q}\bar{s}$  & $\eta_4$ & 115 & $5.8- 8.6$ & 7.2 & $10.2\pm0.3$  & 45.6 & 0.337 & 10.65\\
                    & $\eta_5$ & 120 & $6.2- 9.8$ & 8.0 & $10.3\pm0.3$  & 49.3 & 0.806 &
                    \vspace{5pt} \\
                    & $\eta_1$ & 130 & $7.5- 9.8$ & 8.5 & $11.0\pm0.2$ & 41.4  & 0.391 & \\
                    &$\eta_2$  & 120 & $6.4- 9.8$ & 8.0 & $10.4\pm0.3$ & 49.7  & 0.632 & \\
$bb\bar{s}\bar{s}$  &$\eta_3$ & 115 & $6.3- 9.0$ & 8.0 & $10.2\pm0.3$  & 40.5  & 0.560 & 10.73\\
                    & $\eta_4$ & 120 & $6.2- 8.4$ & 8.0 & $10.4\pm0.3$  & 41.9 & 0.486 & \\
                    & $\eta_5$ & 115 & $6.2- 8.8$ & 8.0 & $10.2\pm0.3$  & 38.9 & 1.14  & \\
\hlinewd{.8pt}
\end{tabular*}
\tabcaption{Numerical results for the doubly-charmed/bottomed
$QQ\bar q\bar q$, $QQ\bar u\bar d$, $QQ\bar q\bar s$ and $QQ\bar
s\bar s$ systems with $J^P=0^+$.} \label{table0+}
\end{center}

\begin{center}
\renewcommand{\arraystretch}{1.3}
\begin{tabular*}{16cm}{lcccccccc}
\hlinewd{.8pt}
  &Current & $s_0$  & [$M^2_{Bmin},M^2_{Bmax}$] & $M^2_B$ & $m_X $ & ~~~~PC~~~~ & ~~~~$f_X$~~~~ & open charm/bottom \\
                    &  & ($\mbox{GeV}^2$) & ($\mbox{GeV}^2$) &($\mbox{GeV}^2$)& $(\mbox{GeV})$ &(\%)&($\mbox{GeV}^5$) &threshold $(\mbox{GeV})$\\
                    \hline
$cc\bar{q}\bar{q}$  &$\eta_1$ & 23 & $3.0-3.6$ & 3.3 &
$4.35\pm0.14$ & 38.6 & 0.0490  &
 \vspace{5pt} \\
$cc\bar{u}\bar{d}$  &$\eta_6$ & 23 & $3.1-3.7$ & 3.4 & $4.34\pm0.16$ & 37.9 & 0.0395  & 3.730\\
                    &$\eta_7$ & 22 & $2.6-3.4$ & 3.0 & $4.41\pm0.12$ & 39.4 & 0.0690  & \\
                    &$\eta_8$ & 23 & $2.6-3.5$ & 3.0 & $4.42\pm0.14$ & 41.1 & 0.0940  &
                    \vspace{5pt} \\
$cc\bar{q}\bar{s}$ &$\eta_1$ & 23 & $2.7-3.6$ & 3.2 & $4.37\pm0.17$ & 39.1 & 0.0357 &\\
                    &$\eta_2$ & 24 & $2.9-3.8$ & 3.2 & $4.59\pm0.13$ & 43.0 & 0.0838 & \\
                    &$\eta_6$ & 23 & $2.9-3.7$ & 3.4 & $4.35\pm0.16$ & 37.7 & 0.0353 & 3.833\\
                    &$\eta_7$ & 24 & $2.4-3.9$ & 3.4 & $4.58\pm0.14$ & 39.8 & 0.105  & \\
                    &$\eta_8$ & 24 & $2.4-3.9$ & 3.4 & $4.52\pm0.13$ & 41.1 & 0.114  &
                    \vspace{5pt} \\
                    &$\eta_1$ & 24 & $2.8-3.7$ & 3.3 & $4.47\pm0.13$  & 40.7 & 0.0603 & \\
$cc\bar{s}\bar{s}$  &$\eta_3$ & 23 & $2.5-3.5$ & 3.0 & $4.47\pm0.14$  & 40.9 & 0.101  & 3.937\\
                    &$\eta_4$ & 26 & $2.8-4.2$ & 3.3 & $4.74\pm0.17$ & 49.1  & 0.196  &
                    \vspace{5pt} \\
$bb\bar{q}\bar{q}$  &$\eta_1$ & 125 & $7.0- 9.6$ & 8.0 &
$10.6\pm0.3$  & 47.8 & 0.229  &
\vspace{5pt} \\
$bb\bar{u}\bar{d}$  &$\eta_6$ & 120 & $7.2- 8.9$ & 8.0 & $10.4\pm0.2$  & 40.5 & 0.142  & 10.56\\
                    &$\eta_8$ & 120 & $8.2- 9.4$ & 8.8 & $10.5\pm0.2$  & 35.9 & 0.492  &
                    \vspace{5pt} \\
                    &$\eta_1$ & 120 & $7.2- 8.8$ & 8.0 & $10.5\pm0.2$  & 37.9 & 0.124 & \\
$bb\bar{q}\bar{s}$  &$\eta_6$ & 120 & $7.2- 8.9$ & 8.0 & $10.4\pm0.2$  & 40.6 & 0.145 & 10.65\\
                    &$\eta_8$ & 120 & $7.6- 9.3$ & 8.4 & $10.5\pm0.2$  & 37.8 & 0.491 &
                    \vspace{5pt} \\
                    & $\eta_1$ & 125 & $6.6- 9.6$ & 8.0 & $10.6\pm0.3$  & 47.1 & 0.240 & \\
$bb\bar{s}\bar{s}$  & $\eta_3$ & 120 & $6.7- 9.0$ & 8.0 & $10.5\pm0.3$  & 40.1 & 0.490 & 10.73\\
                    & $\eta_4$ & 120 & $6.8- 8.9$ & 8.0 & $10.6\pm0.3$  & 38.8 & 0.655 & \\
\hlinewd{.8pt}
\end{tabular*}
\tabcaption{Numerical results for the doubly-charmed/bottomed
$QQ\bar q\bar q$, $QQ\bar u\bar d$, $QQ\bar q\bar s$ and $QQ\bar
s\bar s$ systems with $J^P=1^-$.} \label{table1-}
\end{center}

\begin{center}
\renewcommand{\arraystretch}{1.3}
\begin{tabular*}{16cm}{lcccccccc}
\hlinewd{.8pt}
  &Current & $s_0$  & [$M^2_{Bmin},M^2_{Bmax}$] & $M^2_B$ & $m_X $ & ~~~~PC~~~~ & ~~~~$f_X$~~~~ & open charm/bottom \\
                    &  & ($\mbox{GeV}^2$) & ($\mbox{GeV}^2$) &($\mbox{GeV}^2$)& $(\mbox{GeV})$ &(\%)&($\mbox{GeV}^5$) &threshold $(\mbox{GeV})$\\
                    \hline
                    & $\eta_1$ & 28 & $3.0- 4.2$ & 3.6 & $4.96\pm0.11$  & 42.1  & 0.0801 &\\
                    & $\eta_2$ & 27 & $3.1- 4.0$ & 3.6 & $4.87\pm0.11$  & 38.5  & 0.0726 & \\
$cc\bar{q}\bar{s}$  & $\eta_3$ & 21 & $2.4- 3.4$ & 2.8 & $4.12\pm0.17$  & 47.5  & 0.0571 & \\
                    & $\eta_4$ & 21 & $2.5- 3.4$ & 2.8 & $4.13\pm0.16$  & 47.9  & 0.0574 & 3.975\\
                    & $\eta_5$ & 21 & $2.8-3.7$ & 3.2 & $4.12\pm0.16 $  & 41.7  & 0.0378 & \\
                    & $\eta_6$ & 21 & $3.0-3.7$ & 3.2 & $4.17\pm0.12 $  & 41.5  & 0.0718 & \\
                    & $\eta_7$ & 21 & $2.2-3.3$ & 2.8 & $4.15\pm0.17 $  & 42.9  & 0.0465 &
                    \vspace{5pt} \\
$cc\bar{s}\bar{s}$ & $\eta_1$ & 29  & $3.2-4.5$ & 3.8 & $5.03\pm0.13$  & 42.5  & 0.138  &\\
                    & $\eta_2$ & 30  & $3.2-4.6$ & 3.8 & $5.12\pm0.14$  & 45.9  & 0.150  & 4.081\\
  & $\eta_3$ & 21  & $2.2-3.4$ & 2.8 & $4.17\pm0.16$  & 45.4  & 0.0838 & \\
                    & $\eta_4$ & 21  & $2.2-3.4$ & 2.8 & $4.19\pm0.16$  & 45.7  & 0.0849 &
                    \vspace{5pt} \\
$bb\bar{q}\bar{q}$  & $\eta_3$ & 115 & $6.5- 8.8$ & 7.8 & $10.2\pm0.3$  & 41.4  & 0.459  &\\
                    & $\eta_4$ & 115 & $6.8- 8.8$ & 7.8 & $10.2\pm0.3$  & 41.7  & 0.454  &
                    \vspace{5pt} \\
$bb\bar{u}\bar{d}$  & $\eta_5$ & 115 & $7.0- 9.0$ & 8.0 & $10.2\pm0.3$  & 42.8  & 0.215  & 10.60\\
                    & $\eta_6$ & 115 & $7.0- 9.2$ & 8.0 & $10.2\pm0.3$  & 42.0  & 0.304  & \\
                    & $\eta_7$ & 115 & $6.5- 8.6$ & 7.6 & $10.2\pm0.3$  & 43.2  & 0.241  & \\
                    & $\eta_8$ & 115 & $6.8- 8.8$ & 7.6 & $10.2\pm0.3$  & 41.7  & 0.343  &
                    \vspace{5pt} \\
                    & $\eta_1$ & 125 & $6.9- 8.6$ & 7.6 & $10.7\pm0.3$  & 42.1  & 0.155  &\\
                    & $\eta_2$ & 125 & $6.9- 8.8$ & 7.6 & $10.7\pm0.4$  & 44.5  & 0.170  & \\
$bb\bar{q}\bar{s}$  & $\eta_3$ & 120 & $6.2- 9.8$ & 8.0 & $10.4\pm0.3$  & 48.9  & 0.452  & \\
                    & $\eta_4$ & 120 & $6.5- 9.8$ & 8.0 & $10.4\pm0.3$  & 49.3  & 0.446  & 10.69\\
                    & $\eta_5$ & 120 & $6.6- 9.8$ & 8.0 & $10.3\pm0.3$  & 52.3  & 0.298  & \\
                    & $\eta_6$ & 120 & $6.6- 9.8$ & 8.0 & $10.3\pm0.4$  & 52.1  & 0.418  & \\
                    & $\eta_7$ & 120 & $6.2- 9.6$ & 8.0 & $10.4\pm0.3$  & 48.1  & 0.342  & \\
                    & $\eta_8$ & 120 & $5.8- 9.6$ & 8.0 & $10.4\pm0.3$  & 46.3  & 0.491  &
                    \vspace{5pt} \\
$bb\bar{s}\bar{s}$  & $\eta_1$ & 130 & $7.0- 9.7$ & 8.5 & $11.0\pm0.3$  & 40.8  & 0.336 &\\
                    & $\eta_2$ & 130 & $7.2- 9.9$ & 8.5 & $10.9\pm0.3$  & 42.9  & 0.370 & 10.78\\
                    & $\eta_3$ & 120 & $6.2- 9.8$ & 8.0 & $10.4\pm0.3$  & 48.1  & 0.657 & \\
                    & $\eta_4$ & 120 & $6.2- 9.8$ & 8.0 & $10.4\pm0.3$  & 48.5  & 0.651 & \\
\hlinewd{.8pt}
\end{tabular*}
\tabcaption{Numerical results for the doubly-charmed/bottomed
$QQ\bar q\bar q$, $QQ\bar u\bar d$, $QQ\bar q\bar s$ and $QQ\bar
s\bar s$ systems with $J^P=1^+$.} \label{table1+}
\end{center}

After performing the QCD sum rule analysis to the
doubly-charmed/bottomed $QQ\bar q\bar q$, $QQ\bar u\bar d$, $QQ\bar
q\bar s$ and $QQ\bar s\bar s$ systems with $J^P=0^-, 0^+, 1^-$ and
$1^+$, we collect the numerical results for all these tetraquark
states in Tables \ref{table0-} -- \ref{table1+} \cite{2013-Du-p14003-14003}. 
We take into account only the
uncertainty of the values of the threshold parameter and variation
of the Borel mass to obtain the errors. The other possible error
sources, including the truncation of the OPE series, the uncertainty
of the quark masses and the condensate values, are
not considered.
From these results,
one finds that there are no stable sum rules for the $0^+$ and $1^+$
$cc\bar q\bar q$ and $cc\bar u\bar d$ systems, which implies that
these tetraquark states probably do not exist. The corresponding
$bb\bar q\bar q$ and $bb\bar u\bar d$ sum rules are relatively more
stable.

We also give the open charm/bottom thresholds for all tetraquark
states in Tabels \ref{table0-} -- \ref{table1+}. One can easily find
that the extracted masses of the $cc\bar{q}\bar{q}$,
$cc\bar{q}\bar{s}$, and $cc\bar{s}\bar{s}$ doubly charmed states are
above the $D^{(*)(+/0)}_{(0/1)}D^{(*)(+/0)}_{(0/1)}$,
$D^{(*)+}_{(0/1)}D^{(*)+}_{s(0/1)}$,
$D^{(*)+}_{s(0/1)}D^{(*)+}_{s(0/1)}$, and $\bar N\Omega_{cc}$
thresholds. They can decay into the two meson/baryon final states
easily through the fall-apart mechanism. They are very broad
resonances and difficult to be observed experimentally. However, the
situations are very different for the doubly bottomed systems. In
Tables \ref{table0-} -- \ref{table1+}, the masses of the
$bb\bar{q}\bar{q}$, $bb\bar{u}\bar{d}$, $bb\bar{q}\bar{s}$, and
$bb\bar{s}\bar{s}$ are below the $\bar{B^0}\bar{B^0},
\bar{B^0_s}\bar{B^0_s}$ and $\bar N+\Omega_{bb}$ thresholds. In
other words, the tetraquark states $bb\bar{q}\bar{q}$,
$bb\bar{u}\bar{d}$, $bb\bar{q}\bar{s}$, and $bb\bar{s}\bar{s}$ are
stable i.e, they only decay via electromagnetic and weak interactions.
This observation is consistent with the conclusions in Refs.~\cite{1988-Carlson-p744-744,2008-Zhang-p437-440,1993-Manohar-p17-33}.

 \section{Mass spectrum of the open-flavor tetraquark states}\label{sec:openflavor}
The $bc\bar q\bar q$ systems were studied in 
Refs.~\cite{1993-Silvestre-Brac-p457-470,1986-Zouzou-p457-457,2013-Feng-p-} and 
their mass predictions are below the thresholds of $B^-D^+$ and $\bar B^0D^0$. 
In Refs.~\cite{2009-Zhang-p56004-56004,2012-Sun-p94008-94008,2012-Albuquerque-p492-498}, 
the authors studied the $c\bar q\bar bq$ systems and indicated that there may 
exist $B_c$-like molecular states. In this section we study the open-flavor 
$bc\bar q\bar q$, $bc\bar s\bar s$ and $qc\bar q\bar b$, $sc\bar s\bar b$ thetraquark 
systems with $J^P=0^+$ and $1^+$ in QCD sum rules.

The QCD sum rules analyses are the same with the previous sections and we ignore the 
details here. We collect the numerical results for the scalar and axial-vector 
$bc\bar q\bar q, bc\bar s\bar s$ tetraquarks states in Tables \ref{tablebcqq0+} and 
\ref{tablebcqq1+} while for $qc\bar q\bar b, qc\bar s\bar b$ tetraquarks states in 
Tables \ref{tableqcqb0+} and \ref{tableqcqb1+}, respectively \cite{2013-Chen-p-b}.
From these results, 
we find that the $bc\bar q\bar q, bc\bar s\bar s$ tetraquark systems have much bigger 
pole contributions than the $qc\bar q\bar b, qc\bar s\bar b$ systems, which results 
in broader Borel windows for the previous systems. The mass sum rules for the 
$bc\bar q\bar q, bc\bar s\bar s$ systems are more stable. 

The numerical results in Tables \ref{tablebcqq0+}--\ref{tableqcqb1+} show that the 
$bc\bar{q}\bar{q}$, $bc\bar{s}\bar{s}$ and $qc\bar{q}\bar{b}$, $sc\bar{s}\bar{b}$ 
tetraquark states lie below the open-flavor thresholds $D^{(\ast)}\bar B^{(\ast)}$,
$D_s^{(\ast)}\bar B_s^{(\ast)}$ and $D^{(\ast)}B^{(\ast)}$, $D_s^{(\ast)}B_s^{(\ast)}$, 
respectively. They cannot decay into the open-flavor final states via the strong 
interaction due to the kinematic limits. But the $qc\bar{q}\bar{b}$ and $sc\bar{s}\bar{b}$
states can decay into $B_c$ plus a light meson, such as $X (0^+)\to B_c\pi, B_c\eta$ 
and $X (1^+)\to B_c\rho, B_c\omega$. Such channels are suggested for the future search 
of these possible $qc\bar{q}\bar{b}$, $sc\bar{s}\bar{b}$ states. However, the 
$bc\bar{q}\bar{q}$ and $bc\bar{s}\bar{s}$ tetraquark states cannot decay through 
these fall-apart mechanisms, suggesting dominantly weak decay mechanisms.
\begin{center}
\renewcommand{\arraystretch}{1.3}
\begin{tabular*}{11.3cm}{cccccccc}
\hlinewd{.8pt}
System            & Current  & $s_0 (\mbox{GeV}^2)$ & $[M^2_{\mbox{min}}$,$M^2_{\mbox{max}}] (\mbox{GeV}^2)$ & $m_X$\mbox{(GeV)}&~~~~PC(\%)\\
\hline
$bc\bar q\bar q$   & $J_{1}$   &  $60\pm2$               & $5.4-6.2$                   & $7.27\pm0.19$    & 35.5  \\
                   & $J_{2}$   &  $59\pm2$               & $6.1-6.4$                   & $7.16\pm0.16$    & 32.9  \\
                   & $J_{3}$   &  $58\pm2$               & $5.4-6.0$                   & $7.14\pm0.16$    & 33.9  \\
                   & $J_{4}$   &  $60\pm2$               & $6.1-6.4$                   & $7.23\pm0.19$    & 33.5
\vspace{5pt}\\
$bc\bar s\bar s$   & $J_{1}$   &  $61\pm2$               & $4.9-6.4$                   & $7.35\pm0.17$    & 39.1  \\
                   & $J_{4}$   &  $60\pm2$               & $5.6-6.5$                   & $7.26\pm0.24$    & 36.7   \\
\hlinewd{1.0pt}
\end{tabular*}
\tabcaption{Numerical results for the $bc\bar q\bar q$ and $bc\bar s\bar s$ systems with $J^P=0^+$.\label{tablebcqq0+}}
\end{center}
\begin{center}
\renewcommand{\arraystretch}{1.3}
\begin{tabular*}{11.3cm}{cccccccc}
\hlinewd{.8pt} System  & Current  & $s_0 (\mbox{GeV}^2)$ & $[M^2_{\mbox{min}}$,$M^2_{\mbox{max}}] (\mbox{GeV}^2)$ & $m_X$
\mbox{(GeV)}&~~~~PC(\%)\\
\hline
$bc\bar q\bar q$   & $J_{1\mu}$   &  $59\pm2$               & $5.5-6.1$                  & $7.21\pm0.16$    & 34.7  \\
                   & $J_{2\mu}$   &  $60\pm2$               & $5.3-6.2$                  & $7.27\pm0.20$    & 37.5  \\
                   & $J_{3\mu}$   &  $60\pm2$               & $5.4-6.3$                  & $7.26\pm0.19$    & 36.8  \\
                   & $J_{4\mu}$   &  $58\pm2$               & $5.3-6.0$                  & $7.13\pm0.17$    & 35.7
\vspace{5pt}\\
$bc\bar s\bar s$   & $J_{2\mu}$   &  $61\pm2$               & $4.9-6.4$                  & $7.35\pm0.22$    & 41.2  \\
                   & $J_{3\mu}$   &  $61\pm2$               & $4.9-6.4$                  & $7.34\pm0.22$    & 42.1   \\
\hlinewd{1.0pt}
\end{tabular*}
\tabcaption{Numerical results for the $bc\bar q\bar q$ and $bc\bar s\bar s$ systems with $J^P=1^+$.
\label{tablebcqq1+}}
\end{center}
\begin{center}
\renewcommand{\arraystretch}{1.3}
\begin{tabular*}{11.3cm}{cccccccc}
\hlinewd{.8pt} System            & Current  & $s_0 (\mbox{GeV}^2)$
& $[M^2_{\mbox{min}}$,$M^2_{\mbox{max}}] (\mbox{GeV}^2)$ &
$m_X$\mbox{(GeV)}&~~~~PC(\%)\\
\hline
$qc\bar q\bar b$   & $J_{1}$   &  $55\pm2$               &$7.8-8.0$                     &$7.11\pm0.15$    & 10.2
\vspace{5pt}\\
$sc\bar s\bar b$   & $J_{1}$   &  $56\pm2$               & $6.6-8.1$                    &$7.16\pm0.18$    & 14.4  \\
                   & $J_{2}$   &  $56\pm2$               & $8.8-9.2$                    &$7.10\pm0.26$    & 10.6  \\
                   & $J_{4}$   &  $56\pm2$               & $8.8-9.1$                    &$7.10\pm0.27$    & 10.9   \\
\hlinewd{1.0pt}
\end{tabular*}
\tabcaption{Numerical results for the $qc\bar q\bar b$ and $sc\bar s\bar b$ systems with $J^P=0^+$.
\label{tableqcqb0+}}
\end{center}
\begin{center}
\renewcommand{\arraystretch}{1.3}
\begin{tabular*}{11.3cm}{cccccccc}
\hlinewd{.8pt}
System             & Current  & $s_0 (\mbox{GeV}^2)$ & $[M^2_{\mbox{min}}$,$M^2_{\mbox{max}}] (\mbox{GeV}^2)$ & $m_X$\mbox{(GeV)}&~~~~PC(\%)\\
\hline
$qc\bar q\bar b$   & $J_{1\mu}$   &  $55\pm2$               & $7.9-8.2$                   & $7.10\pm0.16$    & 10.4  \\
                   & $J_{2\mu}$   &  $55\pm2$               & $7.9-8.2$                   & $7.09\pm0.16$    & 10.7
\vspace{5pt}\\
$sc\bar s\bar b$   & $J_{1\mu}$   &  $55\pm2$               & $6.7-7.9$                   & $7.11\pm0.16$    & 14.0  \\
                   & $J_{2\mu}$   &  $56\pm2$               & $6.7-8.3$                   & $7.15\pm0.20$    & 14.2  \\
                   & $J_{3\mu}$   &  $52\pm2$               & $6.7-7.3$                   & $6.90\pm0.14$    & 11.6  \\
                   & $J_{4\mu}$   &  $52\pm2$               & $6.7-7.3$                   & $6.92\pm0.18$    & 11.0  \\
\hlinewd{1.0pt}
\end{tabular*}
\tabcaption{Numerical results for the $qc\bar q\bar b$ and $sc\bar s\bar b$ systems with $J^P=1^+$.
\label{tableqcqb1+}}
\end{center}

 \section{Mass spectrum of the quarkonium hybrid $\bar QGQ$ states}\label{sec:hybrid}
Including only dimension four condensate in the correlation
functions, the charmonium hybrids with $J^{PC}=0^{-+}, 0^{+-},
1^{-+}, 1^{--}$ and $2^{-+}$ were unstable in 
Refs.~\cite{1985-Govaerts-p215-215,
1985-Govaerts-p575-575,1987-Govaerts-p674-674}. To stabilize these
hybrid sum rules, we reinvestigated the two-point correlation
functions and considered also dimension six tri-gluon condensate
\cite{2013-Chen-p19-19}.

As shown in Fig.~\ref{figOPEcc1-+}, the gluon condensate is the
dominant power correction to the charmonium hybrid sum rule in 
Eq.~\eqref{sumrule}. However, the tri-gluon condensate is too large to 
be neglected. By studying the OPE convergence and the pole
contribution, we obtain the suitable working region of the Borel
mass $4.6$ GeV$^2$ $\leq M_B^2 \leq 6.5$ GeV$^2$ with the continuum
threshold $s_0=17$ GeV$^2$. In Fig.~\ref{fig1-+cc}, the Borel curves
are very stable in the the regions of these parameters. We then
extracted the mass of the $1^{-+}$ charmonium hybrid as \cite{2013-Chen-p19-19}
\begin{eqnarray}
m_X=(3.70\pm 0.21) \mbox{GeV},
\end{eqnarray}
which is about $0.5$ GeV lower than the lattice result in 
Ref.~\cite{2012-Liu-p126-126}. One finds that the tri-gluon condensate
can stabilize the hybrid sum rules and lead to the reliable mass
prediction. After performing the sum rule analysis for all channels,
we collect the numerical results for the charmonium and bottomonium
hybrids in Tables \ref{tablecchybrid} and \ref{tablebbhybrid}
respectively \cite{2013-Chen-p19-19}. Only errors from the uncertainties 
in the charm quark mass and the condensates are taken into account. 
We do not consider other possible error sources such as truncation 
of the OPE series, the uncertainty of the threshold value $s_0$ and 
the variation of Borel mass $M_B$.

\begin{center}
\begin{tabular}{c}
\scalebox{0.8}{\includegraphics{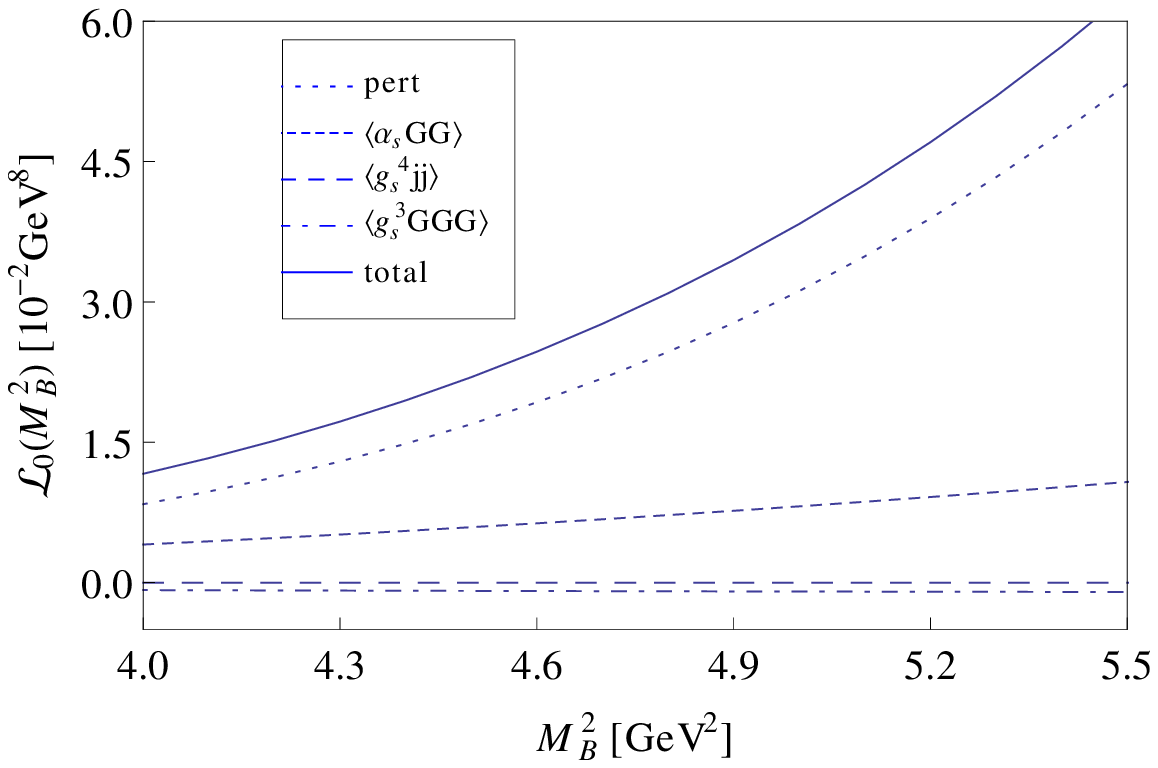}}
\end{tabular}
\figcaption{OPE convergence for the $1^{-+}$ charmonium hybrid.}
\label{figOPEcc1-+}
\end{center}
\begin{center}
\begin{tabular}{lr}
\scalebox{0.7}{\includegraphics{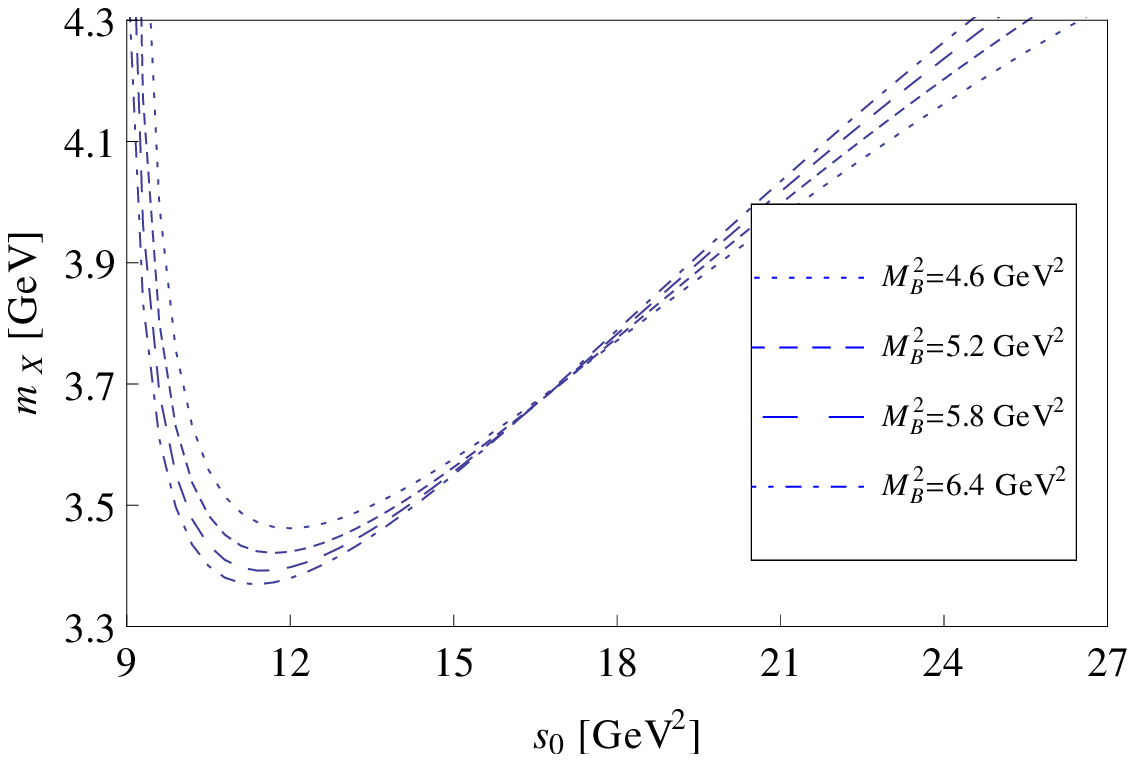}}&
\scalebox{0.7}{\includegraphics{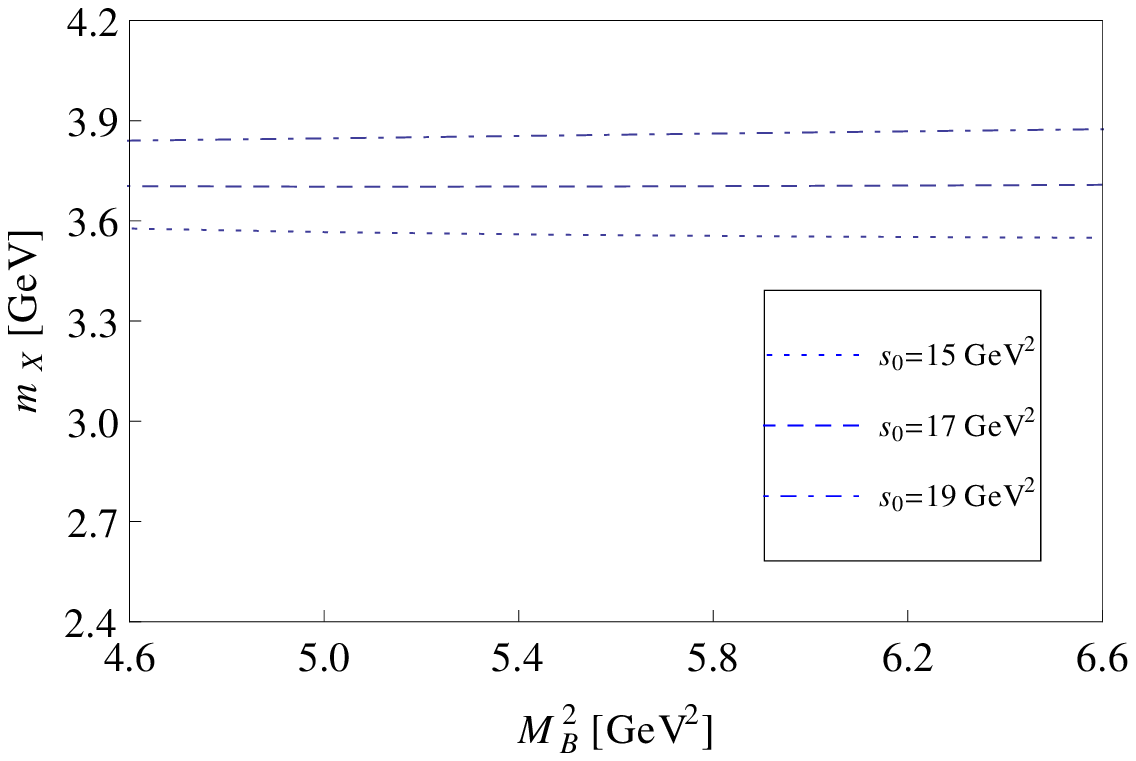}}
\end{tabular}
\figcaption{Variations of $m_X$ with $s_0$ and $M_B^2$ for the
$J^{PC}=1^{-+}$ charmonium hybrid.} \label{fig1-+cc}
\end{center}

\begin{center}
\renewcommand{\arraystretch}{1.3}
\begin{tabular*}{10cm}{lccccc}
\hlinewd{.8pt}
& ~~$J^{PC}$~~ & $s_0(\mbox{GeV}^2)$&$[M^2_{\mbox{min}}$,$M^2_{\mbox{max}}](\mbox{GeV}^2)$&$m_X$\mbox{(GeV)}&~~~~PC(\%)\\
\hline
& $1^{--}$   & 15  & $2.5-4.8 $& $3.36\pm0.15$& 18.3\\
& $0^{-+}$   & 16  & $5.6-7.0 $& $3.61\pm0.21$& 15.4\\
& $1^{-+}$   & 17  & $4.6-6.5 $& $3.70\pm0.21$& 18.8\\
& $2^{-+}$   & 18  & $3.9-7.2 $& $4.04\pm0.23$& 26.0
\vspace{5pt}\\
& $0^{+-}$   & 20  & $6.0-7.4 $& $4.09\pm0.23$& 15.5\\
& $2^{++}$   & 23  & $3.9-7.5$ & $4.45\pm0.27$& 21.5\\
& $1^{+-}$   & 24  & $2.5-8.4 $& $4.53\pm0.23$& 33.2\\
& $1^{++}$   & 30  & $4.6-11.4$& $5.06\pm0.44$& 30.4\\
& $0^{++}$   & 34  & $5.6-14.6$& $5.34\pm0.45$& 36.3
\vspace{5pt}\\
& $0^{--}$    & 35  & $6.0-12.3$& $5.51\pm0.50$& 31.0\\
\hline \hlinewd{.8pt}
\end{tabular*}
\tabcaption{Mass spectrum of the charmonium hybrid states.
\label{tablecchybrid}}
\end{center}

\begin{center}
\renewcommand{\arraystretch}{1.3}
\begin{tabular*}{10cm}{lccccc}
\hlinewd{.8pt}
& ~~$J^{PC}$ ~~& $s_0(\mbox{GeV}^2)$&$[M^2_{\mbox{min}}$,$M^2_{\mbox{max}}](\mbox{GeV}^2)$&$m_X$\mbox{(GeV)}&~~~~PC(\%)\\
\hline
& $1^{--}$    & 105  & $11-17 $& $9.70\pm0.12$&  17.2\\
& $0^{-+}$   & 104  & $14-16 $& $9.68\pm0.29$& 17.3\\
& $1^{-+}$   & 107  & $13-19 $& $9.79\pm0.22$& 20.4\\
& $2^{-+}$   & 105  & $12-19$& $9.93\pm0.21$& 21.7
\vspace{5pt}\\
& $0^{+-}$   & 114  & $14-19 $& $10.17\pm0.22$& 17.6\\
& $2^{++}$   & 120  & $12-20$ & $10.64\pm0.33$& 19.7\\
& $1^{+-}$   & 123  & $10-21 $& $10.70\pm0.53$& 28.5\\
& $1^{++}$  & 134  & $13-27$& $11.09\pm0.60$& 27.7\\
& $0^{++}$  & 137  & $13-31$& $11.20\pm0.48$& 30.0
\vspace{5pt}\\
& $0^{--}$    & 142  & $14-25$& $11.48\pm0.75$& 24.1\\
\hline \hlinewd{.8pt}
\end{tabular*}
\tabcaption{Mass spectrum of the bottomonium hybrid
states.\label{tablebbhybrid}}
\end{center}

$Y(4260)$ was first observed in the $J/\psi\pi^+\pi^-$ channel by
Barbar collaboration \cite{2005-Aubert-p142001-142001} and confirmed
by CLEO \cite{2006-He-p91104-91104} and Belle
\cite{2007-Yuan-p182004-182004} collaborations. Its quantum number
is $J^{PC}=1^{--}$. The open charm decay mode $Y(4260)\to D\bar D$
has not been observed in spite of the large phase space. This is
consistent with the expectation of the hybrid meson decay pattern,
which disfavors the two $S$-wave meson final states but prefers the
$S+P$ decay mode. Since its discovery, $Y(4260)$ was considered as a
good candidate of the charmonium hybrid \cite{2005-Zhu-p212-212,
2005-Kou-p164-169,2005-Close-p215-222}. However, the mass of the
$1^{--}$ channel of charmonium hybrid in Table \ref{tablecchybrid}
is about $3.36\pm 0.15$ GeV, which is much lower than the mass of
$Y(4260)$ meson.

In the MIT bag model
\cite{1983-Barnes-p241-241,1983-Chanowitz-p211-211}, the hybrid
states with $J^{PC}=(0, 1, 2)^{-+}, 1^{--}$ were considered to be
composed of a $S$-wave color-octet $q\bar q$ pair and an excited
gluon field with $J_g^{P_gC_g}=1^{+-}$. This supermultiplet was
confirmed in lattice QCD \cite{2012-Liu-p126-126} and the $P$-wave
quasi gluon approach \cite{2008-Guo-p56003-56003} for the heavy
quarkonium hybrid systems, in which a heavier hybrid supermultiplet
was also predicted including states with $J^{PC}=0^{+-}, (1^{+-})^3,
(2^{+-})^2, 3^{+-}, (0, 1, 2)^{++}$. In Tables \ref{tablecchybrid}
and \ref{tablebbhybrid}, our results support such supermultiplet
structures that the hybrid states with $J^{PC}=(0, 1, 2)^{-+},
1^{--}$ form the lowest supermultiplet while those with the quantum
numbers $J^{PC}=0^{+-}, 2^{++}, 1^{+-}, 1^{++}, 0^{++}$ form a
heavier supermultiplet. The hybrid with $J^{PC}=0^{--}$ is the
heaviest one, which may suggest that this state has a higher gluonic
excitation.

The numerical results of the $\bar bgc$ hybrid states are 
collected in Table \ref{tablebgchybrid} \cite{2014-Chen-p25003-25003}.
In the $\bar bGc$ hybrid 
systems, the supermultiplet structures are still present. 
In Table \ref{tablebgchybrid}, the hybrid states with 
$J^{P}=(0, 1, 2)^{-}, 1^{-}$ form the lightest supermultiplet while 
a heavier one is formed by the states with 
$J^{P}=(0^{+})^2, (1^{+})^2, 2^{+}$. We obtain two 
vector states with  $J^{P}=1^-$ in the lightest hybrid supermultiplet 
while two $1^+$ two $0^+$ hybrids in the heavier supermultiplet. 
The mass differences for the vector, axial-vector and scalar doublet 
are $0.12$ GeV, $0.53$ GeV and $1.18$ GeV, respectively. There also 
exists a pseudoscalar doublet with the mass difference around $1.58$ GeV. 
The existence of these hybrid doublets suggests that the operators 
are separately probing a ground and excited state. The two hybrids 
with the same quantum numbers have very different gluonic excitations. 
\begin{center}
\renewcommand{\arraystretch}{1.3}
\begin{tabular*}{11cm}{ccccccc}
\hlinewd{.8pt}
& Operator & $J^{P}$     & $s_0(\mbox{GeV}^2)$&$[M^2_{\mbox{min}}$,$M^2_{\mbox{max}}](\mbox{GeV}^2)$&$m_X$\mbox{(GeV)}&~~~~PC(\%)\\
\hline
&$\tilde J^{(2)}_{\mu}$   & $1^{-}$   & 52  & $5.90-6.50 $ & $6.83\pm0.16$& 55.9 \\
&$\tilde J^{(1)}_{\mu}$   & $0^{-}$   & 61  & $10.4-11.5 $ & $6.90\pm0.22$& 23.4 \\
&$J^{(1)}_{\mu}$          & $1^{-}$   & 62  & $9.00-10.7 $ & $6.95\pm0.22$& 26.1 \\
&$J^{(3)}_{\mu\nu}$       & $2^{-}$   & 59  & $8.00-10.5 $ & $7.15\pm0.22$& 29.4 
\vspace{5pt}\\
&$\tilde J^{(2)}_{\mu}$   & $0^{+}$   & 69  & $10.9-12.0 $ & $7.37\pm0.31$& 22.8 \\
&$\tilde J^{(3)}_{\mu\nu}$& $2^{+}$   & 66  & $8.00-11.2 $ & $7.67\pm0.18$& 39.8 \\
&$J^{(2)}_{\mu}$          & $1^{+}$   & 71  & $5.90-8.00 $ & $7.77\pm0.24$& 59.4 \\
&$\tilde J^{(1)}_{\mu}$   & $1^{+}$   & 77  & $9.00-10.0 $ & $8.28\pm0.38$& 54.3 \\
&$J^{(1)}_{\mu}$          & $0^{+}$   & 84  & $10.4-12.2 $ & $8.55\pm0.44$& 55.7 
\vspace{5pt}\\
&$J^{(2)}_{\mu}$          & $0^{-}$   & 76  & $10.9-14.2 $ & $8.48\pm0.67$& 24.4 \\
\hline
\hlinewd{.8pt}
\end{tabular*}
\tabcaption{Numerical results of the $\bar bgc$ hybrid states.\label{tablebgchybrid}}
\end{center}

 \section{Summary}\label{sec:sum}
In this article, we have reviewed our previous investigations of the
quarkoniumlike tetraquark $qQ\bar q\bar Q$ systems \cite{2010-Chen-p105018-105018,2011-Chen-p34010-34010}, doubly
charmed/bottomed tetraquark $QQ\bar q\bar q$ systems \cite{2013-Du-p14003-14003}, open-flavor 
$bc\bar q\bar q$ and $qc\bar q\bar b$ systems \cite{2013-Chen-p-b}, heavy
quarkonium hybrid $\bar QGQ$ systems \cite{2013-Chen-p19-19} and the bottom-charm hybrid 
$\bar bGc$ systems \cite{2014-Chen-p25003-25003} in the QCD sum rules approach.

The discovery of the $XYZ$ states is a significant challenge to our
understanding of the QCD hadronic spectrum. To understand the nature of these
new mesons, the formalism of QCD sum rules provides is very useful.
We have evaluated the mass spectra of the heavy tetraquarks and the
quarkonium hybrids in the framework of QCD sum rules. The study of
the charmoniumlike tetraquark state $qc\bar q\bar c$ provides
possible interpretations for several new $XYZ$ states, such as
$Y(4660), X(3872), Z_c(3900)$ and $Z_c(4025)$. The extracted mass of
the $1^{--}$ $qc\bar q\bar c$ tetraquark state is about
$m_X=(4.64\pm0.09)$ GeV, which is consistent with the mass of
$Y(4660)$ meson and may indicate a possible tetraquark
interpretation. The mass of $1^{++}$ channel is about
$m_X=(4.03\pm0.11)$ GeV, which is slightly above the mass of
$X(3872)$. Considering the uncertainties, the tetraquark
interpretation of $X(3872)$ is not excluded. The calculations of the
$1^{++}$ and $1^{+-}$ $qc\bar q\bar c$ channels lead to the lowest
lying ground states around 3.9--4.2 GeV, which may support the
charged states $Z_c(3900)$ and $Z_c(4025)$ as the candidate of the
isovector charmoniumlike tetraquark states with $J^P=1^+$.
Surprisingly, the mass spectrum of the charmonium hybrid states in
Table \ref{tablecchybrid} is much lower than that obtained in
lattice QCD \cite{2012-Liu-p126-126}. For non-exotic $J^{PC}$, 
the effect of  mixing with quarkonium states  may raise these 
mass predictions \cite{2013-Chen-p45027-45027}.

All the $bb\bar{q}\bar{q}$, $bb\bar{u}\bar{d}$, $bb\bar{q}\bar{s}$,
$bb\bar{s}\bar{s}$, $bc\bar{q}\bar{q}$ and $bc\bar{s}\bar{s}$ tetraquark states lie below the
$\bar{B^0}\bar{B^0}, \bar{B^0_s}\bar{B^0_s}$, $\bar
N+\Omega_{bb}$, $D^{(\ast)}\bar B^{(\ast)}$ and
$D_s^{(\ast)}\bar B_s^{(\ast)}$ thresholds. These tetraquark states cannot decay via
strong interaction into both the two meson and two baryon final
states. They should be very narrow because they can decay via
electromagnetic and weak interactions only. These states may be searched for at LHCb and RHIC in the
future, where many  heavy quarks are produced.

\section*{ACKNOWLEDGMENTS}

This project was supported by the the National Natural Science
Foundation of China under Grant NO. 11261130311 and the Natural
Sciences and Engineering Research Council of Canada (NSERC).


\end{document}